\documentclass[journal]{IEEEtran}
\IEEEoverridecommandlockouts
\usepackage{cite}
\usepackage{easybmat}
\usepackage{stfloats}
\usepackage{amsmath,amssymb,amsfonts}
\usepackage{graphicx}
\usepackage{textcomp}
\usepackage{epstopdf}

\usepackage[dvipsnames]{xcolor}
\usepackage{caption}
\usepackage{subcaption}
\usepackage{algorithm}
\usepackage{algorithmic}

\usepackage{setspace}

\DeclareFontFamily{U}{mathx}{\hyphenchar\font45}
\DeclareFontShape{U}{mathx}{m}{n}{
      <5> <6> <7> <8> <9> <10>
      <10.95> <12> <14.4> <17.28> <20.74> <24.88>
      mathx10
      }{}
\DeclareSymbolFont{mathx}{U}{mathx}{m}{n}
\DeclareFontSubstitution{U}{mathx}{m}{n}
\DeclareMathAccent{\widecheck}{0}{mathx}{"71}

\usepackage[letterpaper, left=0.7in, right=0.7in, bottom=1in, top=0.7in]{geometry}

\newtheorem{thm}{Theorem}
\newtheorem{prop}{Proposition}

\newtheorem{ex}{Example}
\newtheorem{cor}{Corollary}
\newtheorem{con}{Conjecture}

\newenvironment{proof}{ \paragraph*{\hspace{-1em}Proof}}{\hfill$\square$}

\newenvironment{skproof}{ \paragraph*{\hspace{-1em}Supporting arguments}}{\hfill$\square$}

\renewcommand{\vec}[1]{\ensuremath{\boldsymbol{#1}}}
\newcommand{\bs}[1]{\ensuremath{\boldsymbol{#1}}}
\newcommand{\be}{\begin{equation}}
\newcommand{\ee}{\end{equation}}
\newcommand{\ba}{\begin{array}}
\newcommand{\ea}{\end{array}}
\newcommand{\bea}{\begin{eqnarray}}
\newcommand{\eea}{\end{eqnarray}}
\newcommand{\A}{\vec{A}}

\newcommand{\Atil}{\widetilde{\boldsymbol{A}}}
\newcommand{\Btil}{\widetilde{\boldsymbol{B}}}

\newcommand{\X}{\vec{X}}    
\newcommand{\W}{\vec{W}}

\renewcommand{\H}{\vec{H}}
\newcommand{\mrm}[1]{\ensuremath{\mathrm{#1}}}

\newcommand{\diag}{\mathrm{diag}} 

\newcommand{\tA}{\widetilde{\A}}

\newcommand{\comment}[1]{}

\newcommand{\MP}{M_\mrm{P}}
\newcommand{\TP}{T_\mrm{P}}
\newcommand{\Mdiff}{\Phi}
\begin{document}

\title{Trade-Offs in Decentralized Multi-Antenna Architectures: Sparse Combining Modules for WAX Decomposition}
\author{Juan Vidal Alegr\'ia, and
        Fredrik Rusek% <-this % stops a space
        \thanks{This paper is built upon previous results presented at the 2022 IEEE ICC conference \cite{icc2022}.}
}

\maketitle

\begin{abstract}
With the increase in the number of antennas at  base stations (BSs), centralized multi-antenna architectures have encountered scalability problems from excessive interconnection bandwidth to the central processing unit (CPU), as well as increased processing complexity. Thus, research efforts have been directed towards finding decentralized receiver architectures where a part of the processing is performed at the antenna end (or close to it). A recent paper put forth an information-lossless trade-off between level of decentralization (inputs to CPU) and decentralized processing complexity (multiplications per antenna). This trade-off was obtained by studying a newly defined matrix decomposition--the WAX decomposition--which is directly related to the information-lossless processing that should to be applied in a general framework to exploit the trade-off. {The general framework consists of three stages: a set of decentralized filters, a linear combining module, and a processing matrix applied at the CPU; these three stages are linear transformations which can be identified with the three constituent matrices of the WAX decomposition. The previous work was unable to provide explicit constructions for linear combining modules which are valid for WAX decomposition, while it remarked the importance of these modules being sparse with 1s and 0s so they could be efficiently implemented using hardware accelerators.} In this work we present a number of constructions, as well as possible variations of them, for effectively defining linear combining modules which can be used in the WAX decomposition. Furthermore, we show how these structures facilitate decentralized calculation of the WAX decomposition for applying information-lossless processing in architectures with an arbitrary level of decentralization.
\end{abstract}


\section{Introduction}
\label{section:intro}
\IEEEPARstart{M}{ulti-antenna} architectures constitute a mature technology which keeps developing to improve wireless communication links. Their main benefits include increased data rates and reliability due to the exploitation of space-division multiplexing and diversity. Current research  on multi-antenna architectures is trending towards scaling up the number of antennas in order to further increase spectral efficiency and spatial resolution. %The exploitation of millimeter-wave spectrum in modern communications \cite{5G} further justifies this increase in the number of antennas to cope with the huge path-losses experienced at these frequencies by focusing the transmitted energy more effectively \cite{mmW}. 
This trend can be seen, e.g., in massive multiple-input multiple-output (MIMO) \cite{marzetta,rusek} and large intelligent surface (LIS) \cite{husha_data}, where massive MIMO considers base stations (BSs) with hundreds of antennas, while LIS goes beyond by considering whole walls of electromagnetically active material.

Several prototypes of massive MIMO have been developed and tested \cite{lumami,argos,bigstation}. In the prototypes from \cite{lumami,bigstation}, centralized processing results in scalability issues due to the increased data-rates between the antennas and the central processing unit (CPU), which scales with the number of antennas. These issues become even more concerning in LIS, where practical deployments are expected to include a number of antennas at least an order of magnitude greater than massive MIMO \cite{jesus_lis}.\footnote{Discrete surfaces approximate continuous ones when sampling is dense enough \cite{husha_data,pizzo_nyq}.} Cell-free massive MIMO\cite{cell_free,sc_cell_free,zhang_prosp,cell-free_wax} is also likely to suffer from scalability issues due to the large number of access points (APs) distributed throughout large geographical areas. Our system model will consider a general multi-antenna architecture which can be generalized to more specific applications, e.g., the ones previously mentioned.

Decentralized pre-processing of the received signals at the antenna end (or nearby) allows to reduce the dimension of the data that needs to be transmitted to a CPU\cite{isit_2019,wiffen_dist_mimo,icassp21}. 
In the recent years, there has been a trend towards considering more decentralized architectures \cite{cavallaro,larsson,wiffen_dist_mimo,icassp21,li_tradeoffs,isit_2019,muris,jesus,vtc, zhang, amiri_mess_pass} in order to cope with scalability issues arising in large-scale multi-antenna architectures. The literature on decentralized massive MIMO includes a number of solutions, ranging from fully-decentralized architectures \cite{jesus,muris,vtc, cavallaro}, where channel state information (CSI) does not have to be available at the CPU, to partially decentralized architectures, where some of the processing tasks are distributed, but neither full \cite{larsson,amiri_mess_pass}, nor partial CSI \cite{isit_2019} is available at the CPU. We can also find decentralized solutions tailored for other large-scale multi-antenna systems such as for cell-free massive MIMO \cite{sc_cell_free,cell-free_dec}, or for extra-large scale MIMO (XL-MIMO) \cite{amiri_mess_pass,amiri_dec}, which can be seen as a system with a number of antennas in the regime of massive MIMO where the antenna array is deployed throughout a large surface such that spatial non-stationarities appear \cite{xl-mimo}. 

In \cite{wax_journal}, an information-lossless trade-off between the number of connections to a CPU and number of multiplications per antenna is presented.\footnote{Information-lossless here means that the mutual information between the post-processed and the user data is equal to the mutual information between the received data and the user data.} To this end, a general framework is considered which can accommodate classical centralized processing architectures, decentralized architectures such as \cite{isit_2019}, as well as a wide range of intermediate architectures. Unlike \cite{li_tradeoffs}, where a system-level trade-off between different decentralized architectures, algorithms, and data precision is studied, \cite{wax_journal} gives a fundamental trade-off between level of decentralization and decentralized processing complexity. The information-lossy regime of said trade-off is considered in \cite{jesus_lis,wiffen_dist_mimo}, while we restrict our work to the information-lossless regime. {Hence, the results from \cite{cavallaro,larsson,wiffen_dist_mimo,li_tradeoffs,muris,jesus,vtc, zhang, amiri_mess_pass,amiri_dec} lie essentially outside the scope of our work since they rely on the usage of linear equalizers which incur information-losses before symbol detection, and/or they focus on the symbol detection problem, which we disregard in this work. Furthermore, most of these works focus on the implementation of solutions as decentralized as possible, while our aim is to understand the trade-offs arising when we can have different levels of decentralization. Thus, we consider the general framework from \cite{wax_journal}, which corresponds to a generic architecture useful in the analysis of the information-lossless regime of decentralized linear equalization. Note that BER is not a suitable metric for judging the results presented in this work,\footnote{{BER can be made arbitrarily small when operating at rates below capacity \cite{inf_th} with marginal loss when considering practical channel coding methods, e.g., LDPC \cite{ldpc}.}} while channel capacity is perfectly achievable under this framework. 

The WAX decomposition, as originally introduced in \cite{wax_journal}, is a matrix decomposition which has direct correspondence with the information-lossless linear processing to be applied in an architecture with an arbitrary level of decentralization.} It thus allows to characterize the information-lossless trade-off between level of decentralization and decentralized processing complexity. The idea is to decompose the channel matrix into the product of a (block-diagonal) decentralized processing matrix $\W$, a linear combining module $\A$, and a CPU processing matrix $\X$. In \cite[Theorem~1]{wax_journal}, the requirements for the existence of the WAX decomposition are only proved for randomly chosen channel matrices and using fixed randomly chosen combining modules $\A$ (for definition of "randomly chosen" see \textit{Notation}). In \cite{cell-free_wax}, the applicability of the WAX decomposition is generalized to sparse channel matrices, showing that channel sparsity can degrade the trade-off given in \cite{wax_journal}. {On the other hand, \cite{wax_journal} remarks the importance of employing a simple sparse combining matrix $\A$ with 1s and 0s, so that it could be efficiently implemented through hardware modules, i.e., generalizing the trivial combining modules from purely decentralized architectures (e.g., the sum module from \cite{isit_2019}) or common centralized architectures (i.e., an identity module).} However, \cite{wax_journal} only presents necessary conditions for an $\A$ to be valid for WAX decomposition.

The current paper is a continuation of the work presented in \cite{wax_journal}, and it further extends the results from \cite{icc2022}. Thus, our aim is to fill some of the gaps from \cite{wax_journal} by presenting a set of constructions for $\A$ which consist of sparse structures of 1s and 0s,\footnote{This condition is slightly relaxed in degenerate cases as will be discussed.} and which can be proved valid for WAX decomposition under different parameter settings. {The proven existence of these constructions strengthens the practicality of the WAX decomposition for the exploitation of the trade-off between level of decentralization and decentralized processing complexity from \cite{wax_journal}.} Furthermore, we exploit the structure of said $\A$ matrices to define a decentralized scheme for computing the information-lossless decentralized filters without the need of aggregating the full CSI at any single point. {We also extend \cite[Theorem~1]{wax_journal} by proving the converse (only if) statement for arbitrary combining modules, thus showing that the information-lossless trade-off studied \cite{wax_journal} is of fundamental nature and it is not possible to operate without loss beyond it.} The list of contributions are summarized next:
\begin{itemize}
    {
    \item We prove that there exists no combining module, $\A$, attaining a less-restrictive information-lossless trade-off than the one obtained in \cite[Theorem~1]{wax_journal}, which was only proved for randomly chosen $\A$.
    }
    \item We present an equivalent formulation of the WAX decomposition which describes the information-lossless regime without the need of taking into account any processing at the CPU. This was already included in \cite{icc2022}.
    \item We present 3 sparse structures for $\A$ and prove their validity for WAX decomposition. Only one of these structures was included in \cite{icc2022}. The new structures allow for more freedom in the exploitation without loss of the achievable trade-off, which corresponds to a novel generalization of the trade-off from \cite{wax_journal} with marginal loss.
    \item We present two transformations for $\A$ that maintain its validity. One of them was included in \cite{icc2022}.
    \item We present a general algorithm to construct a matrix $\A$ that allows for the exploitation of the achievable trade-off for any set of system parameters. Unfortunately, we were unable to formally prove the validity of the $\A$ matrices constructed using said algorithm.
    \item We present a decentralized scheme for computing the information-lossless decentralized filters which generalizes the one included in \cite{icc2022} to the new $\A$ structures presented in this work.
\end{itemize}

The rest of the paper is organized as follows. Section~\ref{section:model} presents the system model and discusses the relevant background from \cite{wax_journal}. Section~\ref{section:eqWAX} presents the main theoretical results, including the converse of \cite[Theorem~1]{wax_journal}, and the equivalent formulation of the WAX decomposition. In Section~\ref{section:validA}, we discuss different ways of constructing a valid combining matrix $\A$. Section~\ref{section:decent} describes the decentralized scheme for computing the decentralized filters considering the valid $\A$ structures. In Section~\ref{section:ex_dis}, we present some examples as well as a discussion of the previous results. Finally, Section~\ref{section:conc} concludes the paper.

\textit{Notation:} In this paper, lowercase, bold lowercase and bold uppercase letters stand for scalars, column vectors and matrices, respectively. When using the mutual information operator, $I(\cdot;\cdot)$, bold uppercase  sub-scripts refer to random vectors instead of their realizations. The operations $(\cdot)^{\text{T}}$, $(\cdot)^*$ and $(\cdot)^{\text{H}}$ denote transpose, conjugate, and conjugate transpose, respectively. The operation $(\cdot)^{\dagger}$ denotes Moore-Penrose inverse. The operation $\mathrm{diag}(\cdot,\ldots,\cdot)$ outputs a block diagonal matrix with the input matrices as the diagonal blocks. $\vec{A} \otimes \vec{B}$ denotes the Kronecker product between matrices $\vec{A}$  and $\vec{B}$. $\mathbf{I}_i$ corresponds to the identity matrix of size $i$, $\boldsymbol{1}_{i\times j}$ denotes the $i \times j$ all-ones matrix, and $\boldsymbol{0}_{i \times j}$ denotes the $i \times j$ all-zeros matrix (absence of one such index indicates that the matrix is square). The notation $[\boldsymbol{A}]_{i:j,\ell:k}$ denotes a matrix formed by rows $i$ to $j$ and columns $\ell$ to $k$ of $\vec{A}$ (as in Python vector notation, absence of one or more indexes indicates that start/end of the included rows or columns corresponds to the first/last row or column of $\boldsymbol{A}$, respectively). In this paper, a randomly chosen matrix corresponds to a realization of a random matrix where any submatrix of it is full-rank with probability 1, e.g., a realization of an independent and identically distributed (IID) Gaussian matrix.

\section{System model}\label{section:model}
\begin{figure}[h]
	\centering
	\includegraphics[scale=0.555]{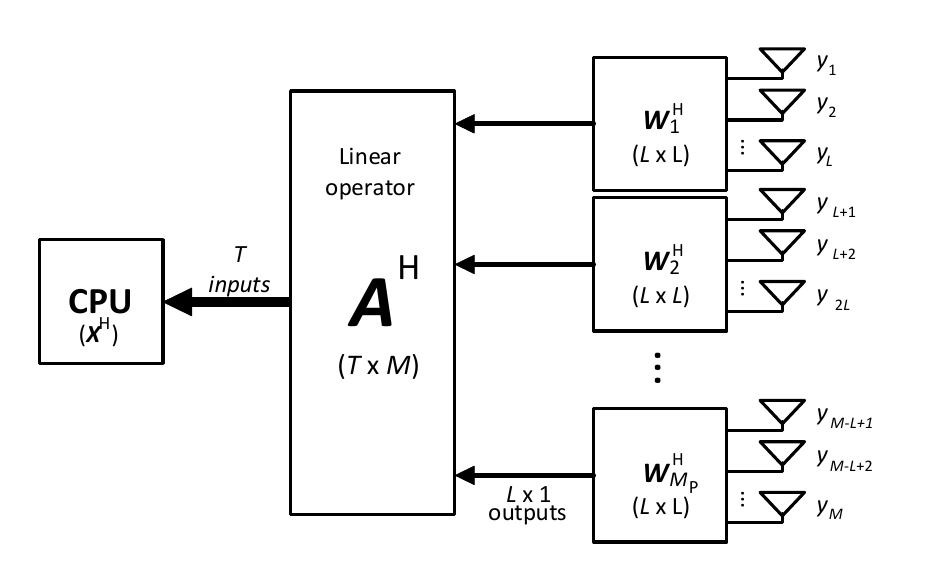}
\caption{Framework considered in this paper during an uplink transmission.}
	\label{fig:gen_arch}
\end{figure}

Let us consider $K$ single-antenna users transmitting to an $M$-antenna BS, with $M>K$, through a narrow-band channel. The $M\times 1$ received complex baseband vector can be expressed as
\begin{equation}
\boldsymbol{y} = \boldsymbol{H}\boldsymbol{s} + \boldsymbol{n},
\label{eq:ul_model}
\end{equation}
where $\boldsymbol{H}$ is the $M\times K$ channel matrix, $\boldsymbol{s}$ is the $K \times 1$ vector of symbols transmitted by the users, and $\boldsymbol{n}$ is a zero-mean complex white Gaussian noise vector with sample variance $N_0$. The $M$ antennas are divided into $\MP$ groups (or panels) with $L$ antennas ($M/L$ evaluates to an integer). Thus, we can express the channel matrix as $\boldsymbol{H}=[\boldsymbol{H}_1^{\text{T}} \, \boldsymbol{H}_2^{\text{T}} \, \dots \boldsymbol{H}_{\MP}^{\text{T}}]^{\text{T}}$ where $\H_m$ corresponds to the $L\times K$ local channel matrix seen by panel $p$, for $p\in \{1,\dots, \MP \}$. Each panel multiplies the received vector by an $L\times L$ matrix, $\boldsymbol{W}_m^{\text{H}}$ $m \in \{ 1,\dots, \MP$\}, thus generating $L$ outputs,\footnote{From \cite{wax_journal}, the restriction of having the same number of antennas and outputs in each panel can be relaxed through an equivalent transformation without constraining the validity of our analysis.} $L\leq K$. The aggregated outputs are combined through a fixed $T\times M$ matrix, $\boldsymbol{A}^{\text{H}}$, $T\leq M$. We can view $\boldsymbol{A}^{\text{H}}$ as a hardware combining module which can be predesigned, but is fixed once deployed. The resulting vector is forwarded to a CPU, which can apply further processing. {In order be able to relate the resulting linear processing to common strategies, e.g., MRC, ZF, MMSE, etc, we assume that the processing at the CPU can be given by a matrix multiplication with a $K\times T$ matrix $\boldsymbol{X}^{\text{H}}$ (see \cite{wax_journal} for further details).} The post-processed vector is then given by
\begin{equation}\label{eq:z_proc}
\boldsymbol{z} = \boldsymbol{X}^{\text{H}} \boldsymbol{A}^{\text{H}} \boldsymbol{W}^{\text{H}} \boldsymbol{y},
\end{equation}
where $\boldsymbol{W}$ is an $M \times M$ block diagonal matrix of the form
\begin{equation}\label{eq:w}
    \W = \diag\left(\W_1,\W_2,\dots,\W_{\MP}\right).
\end{equation}
The matrices $\W$ and $\X$ can be recalculated for every channel realization, while the matrix $\A$ remains unchanged once the system is deployed (we can think of it as a fixed hardware combining module). The framework under study is represented in Fig. \ref{fig:gen_arch}. {Note that, during the whole uplink transmission, information is only flowing from the antennas towards the CPU, unlike message passing approaches like \cite{amiri_mess_pass,amiri_dec,zhang}. This means that there is no extra delay with respect to common centralized architectures, apart from the delay associated to the computation of the decentralized filters which is only done once per coherence interval.}

The main challenge of the current framework is to maximize the information rate at which the users can transmit to the BS, i.e., $I_{\boldsymbol{Z},\boldsymbol{S}}(\boldsymbol{z};\boldsymbol{s})$, or, correspondingly,\footnote{Note that $\boldsymbol{X}$ cannot possibly increase the maximum information rate at which the users can transmit (recall data-processing inequality \cite{inf_th}). The main purpose of it is to be able to consider specific linear receiver schemes, e.g., zero-forzing (ZF), matched filter (MF), etc.} $I_{\boldsymbol{Y},\boldsymbol{S}}(\A^{\text{H}} \W^{\text{H}} \boldsymbol{y};\boldsymbol{s})$. In this paper we will aim at applying information lossless processing, where $I_{\boldsymbol{Y},\boldsymbol{S}}(\A^{\text{H}} \W^{\text{H}} \boldsymbol{y};\boldsymbol{s}) = I_{\boldsymbol{Y},\boldsymbol{S}}(\boldsymbol{y};\boldsymbol{s})$. {Note that the application of $\X$ is not strictly necessary since it cannot possibly increase the information rate.}

The framework under study allows for {an information-lossless} trade-off between the number of multiplications per antenna, $L$, and the number of inputs to the CPU, $T$. Said trade-off was identified in \cite{wax_journal}, where initial results are presented. In the present work we aim at presenting new results that allow for practical exploitation of the trade-off.

Having the number of antennas per panel equal to the multiplications per antenna, both given by $L$ in this work, might seem like unnecessarily restrictive. In \cite{wax_journal}, the number of antennas per panel considered was an arbitrary number $N$, leading to $\W_m$ matrices of size $N \times L$. However, the most important results in said paper consider the case $N=L$ due to its intrinsic generality in the information-lossless scenario. Note that, in order to achieve information-lossless processing, we require $N\leq L$, while if $N$ divides $L$, \cite[Lemma~2]{wax_journal} shows that  there is a direct mapping to the case where $N=L$. Furthermore, from a practical perspective, minimum interconnection bandwidth (i.e., outputs per panel) in the information-lossless case is achieved for $N=L$. Considering all the above, we find it reasonable to focus on the case where the number of antennas per panel coincides with the number of outputs per panel as in the presented framework. However, it would be straightforward to consider panels formed by several of these groups of $L$ antennas as in \cite{cell-free_wax}.

% Maybe this should go in the next section or we should include the previous results section in the system model section?

The framework discussed so far shows how the system operates during the data phase, where the users are transmitting data within one coherence block, so the corresponding $\W$ and $\X$ matrices have already been calculated for the current channel realization $\H$. In this work we also focus on what is being done during the training phase. Specifically, we want to find decentralized schemes to compute the information-lossless decentralized filters to be applied.\footnote{By decentralized here we mean that each panel has access to its local channel, $\H_m$ and it can share some reduced data with a number of other panels to find the processing to be applied.} Since the application of $\X$ at the CPU cannot possibly increase mutual information (as previously discussed), we restrict our problem to proposing a decentralized scheme that allows us to compute the equalizer that each panel has to apply, i.e., $\W_m \; \forall m$, such that the overall processing is information-lossless. In this way, the data arriving at the CPU will contain the same amount of information from the users as in the centralized case. As we will see, the structure of $\A$ plays a big role in how the decentralized computation of $\W$ can be performed. Thus, we will explore how certain structures for $\A$ allow the definition of decentralized schemes for obtaining $\W_m$ at each panel. 
\subsection{Background} 
% Present the main results from previous papers and define the scope of our current paper
As we mentioned earlier, the system model considered in this work was already studied in \cite{wax_journal}, where we can find important results which will be required for our analysis. From \cite[Lemma~1]{wax_journal}, the framework under study can achieve information lossless processing if and only if we can decompose the channel matrix $\H$ into the so called WAX decomposition
\begin{equation}\label{WAX}
    \H = \W \A \X,
\end{equation}
where $\W$, $\A$ and $\X$ correspond to the matrices from \eqref{eq:z_proc}, i.e., $\A$ is fixed by design while $\W$ and $\X$ can be tuned to $\H$. Note that, according to \cite[Lemma~1]{wax_journal}, selecting $\W$ and $\X$ in \eqref{eq:z_proc} such that \eqref{WAX} is fulfilled leads to information-lossless processing within our framework. The main result of the applicability of WAX decomposition is given in \cite[Theorem~1]{wax_journal}, which states that, for a fixed randomly chosen $\boldsymbol{A}\in\mathbb{C}^{M\times T}$, a randomly chosen $\boldsymbol{H}\in\mathbb{C}^{M\times K}$ admits WAX decomposition with probability 1 if
\begin{equation}\label{eq:cond_wax}
T>\max\left(M\frac{K-L}{K},K-1\right).
\end{equation} 
An alternative formulation of \eqref{eq:cond_wax}, can be given by considering the restriction on the other trade-off parameter, $L$. This results in
\begin{equation}
L>K\frac{M-T}{M},
\end{equation}
{ where we restrict ourselves to the regime $T\geq K$ where there exists an information-lossless trade-off between $T$ and $L$ (for $T<K$ there would be information-loss no matter the value of $L$).}
Defining $\TP=T/L$  we have
\begin{equation}\label{eq:cond_wax_L}
L>K\frac{\MP-\TP}{\MP},
\end{equation}
which may ease comparison with the results to be presented.

In this paper, however, we will explore specific structures for $\A$ matrices and prove their validity for WAX decomposition. We consider the same definition as in \cite[Definition~1]{wax_journal} for the validity of $\A$, i.e., a randomly chosen $\H$ admits WAX decomposition with probability 1 for a valid $\A$. Note that \cite{wax_journal} only provides necessary conditions for valid $\A$ matrices which are not randomly chosen, as well as a method to test if a specific $\A$ matrix is valid for some fixed dimensions (not generalizable). It is one of our desires to find structures for $\A$ that allow for a trade-off between $L$ and $T$ as close as possible to \eqref{eq:cond_wax_L} (for $T\geq K$).
\section{New results on the WAX decomposition}
\label{section:eqWAX}
\subsection{The necessary information-lossless trade-off}
{In\cite[Theorem~1]{wax_journal}, the condition \eqref{eq:cond_wax} for the existence of WAX decomposition was only proved for a randomly chosen $\A$. However, it is unclear if there exist any other selection of $\A$ that may attain a better trade-off than the one defined in \eqref{eq:cond_wax}. The following theorem shows that \eqref{eq:cond_wax} is not only a sufficient condition for the existence of the WAX decomposition, but also a necessary condition.
\begin{thm}\label{thm:wax_conv}
Let  $\A$ be an arbitrary $M \times T$,  and $\H$ be an $M \times K$ randomly chosen matrix. The WAX decomposition of $\H$, given by \eqref{eq:cond_wax}, can only exists if \eqref{eq:cond_wax} is satisfied. Furthermore, $\A$ should be of rank $T$ to be able to attain \eqref{eq:cond_wax}.
\begin{proof}
 See Appendix~A.  
\end{proof}
\end{thm}
Theorem~\ref{thm:wax_conv} states that the fundamental trade-off between the number of multiplications per antenna ($L$), and the number of inputs to the CPU ($T$), is ultimately governed by \eqref{eq:cond_wax} (or its alternative formulations). For the rest of the paper we assume $M\geq T\geq K$, which is the regime where the information-lossless trade-off between $L$ and $T$ applies.
}
\subsection{The equivalent formulation of the WAX decomposition}
Let us divide $\A$ into two blocks
\begin{equation}\label{eq:AT_AB}
    \A = \begin{bmatrix} \A_\mathrm{T} \\ \A_\mathrm{B} \end{bmatrix},
\end{equation}
where $\A_\mathrm{T}$ is a $T\times T$  matrix corresponding to the top part of $\A$, and $\A_\mathrm{B}$ is the $(M-T)\times T$ matrix corresponding to the bottom part of $\A$. We next provide a theorem corresponding to an equivalent formulation of the WAX decomposition. 

\begin{thm}\label{thm:AT_AB}
Assume that $\TP=T/L$ evaluates to an integer value, and that $\A_\mathrm{T}$ is full-rank. Then, the WAX decomposition of some $M\times K$ matrix $\H$, given by \eqref{WAX}, exists if and only if we can find a full-rank $\W$ (corresponding to \eqref{eq:w}) such  that
\begin{equation}\label{eq:WAX_eq}
   \boldsymbol{B}^\text{T} \W^{-1}\H=\boldsymbol{0}_{(M-T)\times K},
\end{equation}
where the matrix $\vec{B}$ is defined as
$$
\vec{B} = \begin{bmatrix}\A_\text{B}\A_\text{T}^{-1} & -\mathbf{I}_{M-T} \end{bmatrix}^\text{T}.
$$
%where $\W$ is still of the form \eqref{eq:w}, and it can also be used for solving \eqref{WAX}.
\begin{proof}
Let us assume $\W$ in \eqref{WAX} to be full-rank; correspondingly, $\W_m$ are also full-rank $\forall m$. Note that, considering \cite[Lemma~3]{wax_journal}, the WAX decomposition of a randomly chosen $\H$ exists if and only if then there exists a full-rank $\W$ that achieves said decomposition. From \eqref{WAX} we can get
\begin{equation}\label{eq:X}
    \X = \A_{\mathrm{T}}^{-1}  \diag\left(\W_1,\W_2,\dots,\W_{\TP} \right)^{-1}\begin{bmatrix}\H_1\\\H_2\\ \vdots \\ \H_{\TP}\end{bmatrix},
\end{equation}
where $\A_\mathrm{T}$ is full-rank by assumption. On the other hand, selecting $\X$ as in \eqref{eq:X} implies that, in order to fulfill \eqref{WAX}, we only need to fulfill
\begin{equation}\label{eq:AB_cond}
    \W \A_\text{B}\X=\begin{bmatrix}\H_{\TP+1}\\\H_{\TP+2}\\ \vdots \\ \H_{\MP}\end{bmatrix}.
\end{equation}
If we substitute \eqref{eq:X} in \eqref{eq:AB_cond} and do some simple matrix manipulations we get \eqref{eq:WAX_eq}.
\end{proof}
\end{thm}

We should note that the assumptions taken in Theorem~\ref{thm:AT_AB} are not as restrictive as they seem. In fact, they are fairly reasonable within our framework:
\begin{itemize}
    \item {If $L$ is small with respect to $T$, restricting $\TP=T/L$ to integers will only have a minor effect on the achievable optimum trade-off between $T$ and $L$ from \eqref{eq:cond_wax}. For arbitrary $T$ this restriction can translate to an increase of $T-L\lfloor T/L \rfloor<L$ CPU inputs. Furthermore, in the optimum trade-off regime we have $0\leq L\leq K$ and $K\leq T\leq M$, so $L$ is small with respect to $T$ throughout much of the trade-off, specially as $M$ grows large.}
    \item Having full-rank $\A_\text{T}$ can be relaxed by re-indexing the diagonal blocks of $\W$, i.e., only a submatrix formed by $\TP$ out of the $\MP$ horizontal blocks of dimensions $L\times T$ in $\A$ should be full-rank. Moreover, from Theorem~1, $\A$ should be of rank $T$ to attain \eqref{eq:cond_wax}, so a $T\times T$ submatrix of it should be full-rank.% which, together with \cite[Lemma~4]{wax_journal}, means that the previous requirement is necessary for any valid $\A$.
\end{itemize}
Therefore, we will keep these assumptions throughout the rest of the paper.

The importance of Theorem~\ref{thm:AT_AB} resides in the fact that it provides an alternative formulation of the WAX decomposition without any need to consider $\X$. Since the WAX decomposition allows for information-lossless processing within the framework under study (see \cite{wax_journal}), the new formulation, given in \eqref{eq:WAX_eq}, will also assure information-lossless processing. Thus, we can see \eqref{eq:WAX_eq} as the restriction on the $\W_m$ matrices $\forall m$ in order to achieve information-lossless processing until the CPU is reached (under the assumptions of Theorem~\ref{thm:AT_AB}).

Another important implication of Theorem~\ref{thm:AT_AB} is that we can construct a valid $\A$ matrix by selecting $\A_\text{B}\A_\text{T}^{-1}$ such that there exists a $\W$ that satisfies \eqref{eq:AB_cond} for any randomly chosen $\H$ (except those in a zero-probability set). We can thus note that we have full freedom in selecting $\A_\text{T}$ (as long as it is full-rank) since this matrix can be compensated through a full-rank transformation on $\A_\text{B}$.

Throughout the rest of the paper, we will focus on the study of $\A$ matrices formed as
\begin{equation}\label{eq:A_kron}
    \A = \tA \otimes \mathbf{I}_L,
\end{equation}
where $\tA$ is now an $\MP\times \TP$ matrix. {Even though it may seem like an unnecessary restriction, \eqref{eq:A_kron} is in fact a desirable construction for a number of reasons:
\begin{itemize}
    \item Any $\A$ matrix resulting from \eqref{eq:A_kron} will be inherently sparse since it would have a minimum of $(M-\MP)T$ zeros out of its $MT$ elements.
    \item The combining module resulting from \eqref{eq:A_kron} has a simple hardware implementation since it only requires to scale and phase-shift the aggregated output of each panel before combining it with other panels. In fact, our goal is to eliminate the scaling and phase-shifting so that only sum modules are required.
    \item The equivalent formulation of the WAX decomposition \eqref{eq:WAX_eq} can simplify greatly through \eqref{eq:A_kron}, as will be apparent in Corollary~\ref{cor:Akron}. Hence, it will lead to increased mathematical tractability, allowing to prove the validity of some interesting $\A$ structures.
\end{itemize}}{The main concern that can raise from fixing \eqref{eq:A_kron} is that we may sacrifice achievability of the optimum trade-off \eqref{eq:cond_wax_L}, which is defined for randomly chosen $\A$. However, if we are able to reach a bound arbitrarily close to \eqref{eq:cond_wax_L} we could conclude that there is no loss associated to \eqref{eq:A_kron}.} 

Given \eqref{eq:A_kron}, it becomes natural to extend the definition from \cite[Definition~1]{wax_journal} and talk about valid $\tA$ matrices for WAX decomposition. Considering \eqref{eq:AT_AB}, we can now write
\begin{equation}\label{eq:tAT_tAD}
\begin{aligned}
    \A_\text{T}=\widetilde{\A}_\text{T}\otimes \boldsymbol{I}_L,\\
    \A_\text{B}=\widetilde{\A}_\text{B}\otimes \boldsymbol{I}_L,
\end{aligned}
\end{equation}
where $\widetilde{\A}_\text{T}$ and $\widetilde{\A}_\text{B}$ are matrices of dimensions $\TP\times\TP$ and $(\MP-\TP) \times \TP$, respectively. In order to simplify upcoming notation, let us define
\begin{equation} \label{def_of_Mdiff}
    \Mdiff = \MP-\TP.
\end{equation}
The next corollary comes as a direct consequence of Theorem~\ref{thm:AT_AB} whenever we have \eqref{eq:A_kron}.

\begin{cor}\label{cor:Akron}
Assume that $\A$ is of the form \eqref{eq:A_kron}, and that $\Atil_\text{T}$ is full rank. If we define the matrix 
\begin{equation}\label{eq:B_def}
\Btil = \begin{bmatrix} \widetilde{\A}_\text{B}\widetilde{\A}_\text{T}^{-1} & -\mathbf{I}_{\MP-\TP}\end{bmatrix}^\mathrm{T},
\end{equation}
the WAX decomposition of some $M\times K$ matrix $\H$, given by \eqref{WAX}, exists if and only if we can find full-rank $\W_m$ matrices such that
\begin{equation}\label{eq:WAX_IL}
    \begin{bmatrix}\W_1^{-1}\! & \!\W_2^{-1} \!&\! \dots \! &  \! \W_{\MP}^{-1} \end{bmatrix} \! \begin{bmatrix} \tilde{\vec{b}}_1^\mrm{T} \otimes \H_1 \\ \tilde{\vec{b}}_2^\mrm{T} \otimes \H_2 \\ \vdots\\ \tilde{\vec{b}}_{\MP}^\mrm{T} \otimes \H_{\MP}\end{bmatrix}\!=\boldsymbol{0}_{L\times K\Mdiff},
\end{equation} 
where $\tilde{\vec{b}}_m^\mrm{T}$, for $m = 1,\dots,\MP$, correspond to the rows of $\Btil$. A more compact notation for \eqref{eq:WAX_IL} is achieved by considering the face-splitting product, $(\cdot)\bullet(\cdot)$, which corresponds to a special case of the Khatri-Rao product dividing the left matrix into its rows, i.e.,
\begin{equation}\label{eq:WAX_IL_face}
    \begin{bmatrix}\W_1^{-1} & \W_2^{-1} & \dots & \W_{\MP}^{-1}\end{bmatrix} \left( \Btil \bullet \H \right)=\boldsymbol{0}_{L\times K\Mdiff}.
\end{equation}
\begin{proof}
Let us take Theorem~\ref{thm:AT_AB} and substitute \eqref{eq:tAT_tAD} in \eqref{eq:WAX_eq}. Simple matrix manipulation leads to \eqref{eq:WAX_IL}.
\end{proof}
\end{cor}

Corollary~\ref{cor:Akron} provides a new formulation of the WAX decomposition, now taking into account \eqref{eq:A_kron}. The main benefit of this new formulation is that the diagonal blocks of $\W^{-1}$ come in the form of a block row matrix instead of a block diagonal matrix, which will simplify the tasks of proving valid $\tA$ structures. As happened for $\A$, we can note that the validity of $\tA$ for WAX decomposition depends only on $\Btil$, i.e., the product $\tA_\mathrm{B}\tA_\mathrm{T}^{-1}$ will determine the validity of $\tA$. Our next goal is to come up with clever ways of constructing the product $\tA_\mathrm{B}\tA_\mathrm{T}^{-1}$ which can lead to valid $\tA$.

\section{Constructing valid $\Atil$ matrices}
\label{section:validA}

\subsection{Transforming $\tA$ while maintaining validity}
Taking into account the results from the previous section, we will start by stating some transformations on $\tA$ that maintain its validity for WAX decomposition. These may be useful for proving the validity of specific constructions for $\tA$, {or for generating new $\tA$ structures from those that can be proved valid.}

\begin{prop}\label{prop:inv_trans_A}
Assume a valid $\tA$ for WAX decomposition. If we construct $\tA^\prime = \tA \vec{\Theta}$, where $\vec{\Theta}$ can be any $\TP \times \TP$ full-rank matrix, $\tA^\prime$ is also valid for WAX decomposition.
\begin{proof}
Considering \eqref{eq:B_def} we have that
$$
\begin{aligned}
\Btil^\prime &= \begin{bmatrix} \widetilde{\A}_\text{B}^\prime \widetilde{\A}_\text{T}^{\prime}{}^{-1
} & -\mathbf{I}_{\Mdiff}\end{bmatrix}^\mathrm{T} \\
&= \begin{bmatrix} \widetilde{\A}_\text{B}\vec{\Theta} \vec{\Theta}^{-1}\widetilde{\A}_\text{T}^{-1} & -\mathbf{I}_{\MP-\TP}\end{bmatrix}^\mathrm{T},\\
&=\Btil.
\end{aligned}$$ From Corollary~\ref{cor:Akron} the validity of $\Atil^\prime$ is only determined by $\Btil^\prime$, which leads to Proposition~\ref{prop:inv_trans_A}.
\end{proof}
\end{prop}

The previous proposition can be also trivially extended to $\A$ if we disregard the restriction \eqref{eq:A_kron}. This proposition also remarks that the selection of $\tA_\text{T}$ does not affect the validity of $\tA$ as long as it is full-rank, since it can be compensated by selecting $\vec{\Theta}$.

\begin{prop}\label{prop:perm_A}
Assume $\tA$ is valid for WAX decomposition. If we construct $\tA^\prime = \vec{P}\tA$, where $\vec{P}$ can be any $\MP \times \MP$ permutation matrix, $\tA^\prime$ is also valid for WAX decomposition.
\begin{proof}
It is enough to notice that applying a permutation matrix on $\tA$ only corresponds to a re-indexing of the $\W_m$ matrices in \eqref{eq:w}, which does not affect the solvability of \eqref{WAX}.
\end{proof}
\end{prop}

The previous propositions focused on applying transformations on $\Atil$ that maintain its validity for WAX decomposition. However, as we will see, one way to explore valid $\Atil$ matrices is to explore $\Btil$ matrices of the form \eqref{eq:B_def} that allow us to solve \eqref{eq:WAX_IL_face}. Thus, let us define valid $\Btil$ for WAX decomposition as such matrices allowing for a solution to \eqref{eq:WAX_IL}, i.e., leading to a valid $\A$ through \eqref{eq:tAT_tAD} and \eqref{eq:B_def}.

\subsection{Constructing $\widetilde{\boldsymbol{A}}$ from predesigned $\Btil$}
In Section \ref{section:eqWAX} we noted that properties of $\Btil$, given by \eqref{eq:B_def}, determine the validity of a matrix $\Atil$. We can thus construct an $\Atil$ by first specifying a valid $\Btil$ and then extracting an underlying $\Atil$. More specifically, we should only define the product $\Atil_\text{B} \Atil_\text{T}^{-1}$ giving a valid $\Btil$, and then we can extract a valid $\Atil$ from the possible $\Atil_\text{B}$ and $\Atil_\text{T}$.

If we consider the $\Mdiff\times\TP$ upper part of $\Btil$, given by $(\Atil_\text{B} \Atil_\text{T}^{-1})^\text{T}$, we can note that we have no loss of generality if we set 
\begin{equation}\label{eq:At_Id}
    \tA_\mathrm{T} = \mathbf{I}_{\TP},
\end{equation}
since we can still generate any possible $\Btil$ {of the form \eqref{eq:B_def} by choosing $\Atil_\text{B}$}. \footnote{{Note that with \eqref{eq:At_Id}, $\Atil_\text{B}$ would directly correspond the top $\TP$ rows of $\Btil$, which are the only ones that can be changed for Corollary \ref{cor:Akron} to apply.}}  {Any other full-rank $\Atil_\text{T}$ can be selected by considering the transformation in Proposition~\ref{prop:inv_trans_A}, although said transformation would also change $\widetilde{\A}_\text{B}$. On the other hand, the physical implication of having \eqref{eq:At_Id} is also practically desirable, since this would result in an $\Atil$ with minimum number of 1s in its first $\TP$ rows, i.e., it corresponds to the sparsest possible $\Atil_\text{T}$. The reason is that such $\Atil_\text{T}$ leads, through \eqref{eq:A_kron}, to an $\A$ matrix with a single 1 per row in its first $T$ rows, thus attaining the lower bound from \cite[Lemma~6]{wax_journal}, which corresponds to a lower bound on the number of 1s per row of $\A$ for it to be valid. Therefore, in what follows, we consider $\tA$ matrices such that \eqref{eq:At_Id} is fulfilled. We remark that such selection does not impact the validity of $\tA$ since if we can find a valid $\tA$ with a different $\tA_\mathrm{T}$, we can always find a valid $\tA^\prime$ with $\tA_\mathrm{T}^\prime = \mathbf{I}_{\TP}$ by invoking Proposition~\ref{prop:inv_trans_A} with $\boldsymbol{\Theta}=\tA_\mathrm{T}^{-1}$. Thus, \eqref{eq:At_Id} should not be seen as a restriction, but a beneficial selection of $\tA_\mathrm{T}$ achieving maximum sparsity without loss.}

The following proposition presents a structure for $\A$, taking into account the previous assumptions, which is proved to be valid for WAX decomposition.

\begin{prop}\label{prop:Ab1}
Assume that $\A$ is given by \eqref{eq:A_kron}, with $\tA_\mathrm{T}=\mathbf{I}_{\TP}$, and $\tA_\mathrm{B}$  constructed as
\begin{equation}\label{Ab1}
\begingroup
\renewcommand*{\arraystretch}{1.5}
    \tA_\mathrm{B}=\left[\begin{matrix}
         \boldsymbol{1}_{\Mdiff \times 1} & \boldsymbol{0}_{\Mdiff\times J} & \underbrace{\begin{matrix}
         \boldsymbol{I}_{\Mdiff} &
         \cdots &
         \boldsymbol{I}_{\Mdiff} \end{matrix}
           }_{Q_1=\left\lfloor\frac{\TP-1}{\Mdiff}\right\rfloor}
    \end{matrix}\right],
     \endgroup
\end{equation}
where $J=\TP-1-Q_1\Mdiff$, and where
\begin{equation} \label{def_of_Q1}
    Q_1 = \left\lfloor \frac{\TP-1}{\Mdiff} \right\rfloor.
\end{equation}
A randomly chosen matrix $\H$ admits WAX decomposition with probability 1 for the given $\A$ if
\begin{equation}\label{eq:cond_Ab1}
L \geq \frac{K}{1+Q_1},
\end{equation}
{Furthermore, $\W_1$ (defined in \eqref{eq:w}) can be fixed to an arbitrary $L\times L$ full-rank matrix without affecting the solvability of the WAX decomposition.}
\begin{proof}
Selecting $\tA_\mathrm{T}=\mathbf{I}_{\TP}$ and $\tA_\mathrm{B}$ as in \eqref{Ab1} leads to
$$
\Btil = \left[\begin{matrix}
         \boldsymbol{1}_{\Mdiff \times 1} & \boldsymbol{0}_{\Mdiff\times J} & \underbrace{\begin{matrix}
         \boldsymbol{I}_{\Mdiff} &
         \cdots &
         \boldsymbol{I}_{\Mdiff} \end{matrix}
           }_{Q_1} & -\mathbf{I}_{\Mdiff}
    \end{matrix}\right]^\mathrm{T}.
$$
From Corollary~\ref{cor:Akron}, we can solve the equivalent formulation of the WAX decomposition, given in \eqref{eq:WAX_IL}, with the restriction of having full-rank $\W_m \; \forall m$. If we invoke  \eqref{eq:WAX_IL}, we get the set of equations
\begin{equation} \label{eq:system_Ab1}
\boldsymbol{W}_1^{-1} \boldsymbol{H}_1 = \sum_{q=0}^{Q_1+1} \boldsymbol{W}_{J+r+q\Mdiff}^{-1}\boldsymbol{H}_{J+r+q\Mdiff},\;\; r=1,\dots,\Mdiff.
\end{equation}
Note that we have ignored the negative sign associated to the last identity block in $\Btil$ since this can be absorbed without loss of generality by the corresponding $\H_m$ blocks. {Let us consider $\W_1$ to be fixed to an arbitrary $L\times L$ full-rank matrix (e.g., $\W_1=\mathbf{I}_L$), since this is the only $\W_m$ shared in all the $\Mdiff$ equations from \eqref{eq:system_Ab1}. Note that the selection of $\W_1$, as long as it is full-rank, does not affect the solvability of \eqref{eq:system_Ab1} because this matrix can be absorbed by $\H_1$ (or by the rest of the $\W_m$ matrices) without changing its nature.} Then, through trivial linear algebra arguments, namely counting equations and variables in the resulting linear system, and assuming randomly chosen $\H$ (i.e., $\H_m$ are also randomly chosen $\forall m$ and their sum will reduce rank with probability 0), we can independently solve each of the $\Mdiff$ equations whenever \eqref{eq:cond_Ab1} is fulfilled.
\end{proof}
\end{prop}

The trade-off between $\TP$ and $L$ given by \eqref{eq:cond_Ab1} can be linked to the optimum trade-off for randomly chosen $\A$, given in \eqref{eq:cond_wax_L}, by assuming that $\Mdiff = \MP-\TP$ divides $\TP-1$. In this case we would have,
\begin{equation}\label{eq:new_tradeoff}
    L\geq K\frac{(\MP-\TP)}{\MP-1},
\end{equation}
which for $\MP\gg 1$ corresponds approximately to the same bound as in \eqref{eq:cond_wax_L} {(for small $\MP$, the gap can be linked to the loss of degrees of freedom when fixing $\boldsymbol{W}_1^{-1}$). We can thus conclude that there is essentially no loss in restricting \eqref{eq:A_kron}.} Note that, unlike \eqref{eq:new_tradeoff}, the optimum trade-off \eqref{eq:cond_wax_L} cannot be achieved with equality, which further promotes the equivalence between \eqref{eq:cond_wax_L} and \eqref{eq:new_tradeoff}. Furthermore, due to the integer nature of the variables under consideration, in most cases, both trade-offs would give the same effective parameter restrictions. Let us thus refer to \eqref{eq:new_tradeoff} as the achievable trade-off. The achievable trade-off results from fixing one of the diagonal blocks of $\W$ to identity, as in the proof of Proposition~\ref{prop:Ab1}. 
 
 The main restriction of the construction for $\A$ considered in Proposition~\ref{prop:Ab1} is that the only meaningful points of the achievable trade-off \eqref{eq:cond_Ab1} are those where $\Mdiff$ divides $\TP-1$, since except for those points, there would be an increase in the number of inputs to the CPU, given by $T=L\TP$, without a corresponding decrease in the multiplications per antenna, given by $L$. This restriction becomes specially concerning when we have $\TP< \MP/2+1$, since in this regime Proposition~\ref{prop:Ab1} cannot exploit any trade-off between $T$ (or $\TP$) and $L$. Thus, the following proposition considers a novel structure for $\A$ that allows for exploitation of the trade-off between $T$ and $L$ in the regime $\TP<\MP/2+1$. 

%, but instead of finding a null-space as in \cite{wax_journal}[Theorem~1] we now only need to solve a non-homogeneous system of equations
\begin{prop}\label{prop:Ab2}
Let $\A$ be given by \eqref{eq:A_kron}, with $\tA_\mathrm{T}=\mathbf{I}_{\TP}$, and $\tA_\mathrm{B}$ constructed as
\begin{equation}\label{eq:Ab2}
    \tA_\mathrm{B} = \begin{bmatrix}\alpha_1 \vec{1}_{(\TP-1) \times 1} & \mathbf{I}_{\TP-1}\\
    \vdots & \vdots \\
    \alpha_{Q_2-1} \vec{1}_{(\TP-1)\times 1} & \mathbf{I}_{\TP-1}\\
    \alpha_{Q_2} \vec{1}_{\Pi\times 1} & \left[\mathbf{I}_{\TP-1}\right]_{1:\Pi,:}
    \end{bmatrix},
\end{equation}
where $\Pi=\Mdiff-(Q_2-1)(\TP-1)$, i.e., the last column block is cropped to fit the dimensions, and where
\begin{equation}\label{eq:Q2}
    Q_2 = \left\lceil \frac{\Mdiff}{\TP-1} \right\rceil.
\end{equation}
Furthermore, $\alpha_i\in \mathbb{C} \backslash \{ 0 \}$ can be arbitrarily selected as long as
$$\alpha_i=\alpha_j \iff i=j, \; \forall i,j \in \{1,\dots, Q_2\}.$$
A randomly chosen matrix $\H$ admits WAX decomposition with probability 1 for the given $\A$ if
\begin{equation}\label{eq:cond_Ab2}
L \geq \frac{K}{1+\frac{1}{Q_2}}.
\end{equation}
{Moreover, $\W_1$ (defined in \eqref{eq:w}) can be fixed to an arbitrary $L\times L$ full-rank matrix without affecting the solvability of the WAX decomposition.}
\begin{proof}
See Appendix~B
\end{proof}
\end{prop}
If we assume values of $\MP$ and $\TP$ such that \eqref{eq:Q2} gives an integer without the need of the ceiling operator, the trade-off in \eqref{eq:cond_Ab2} leads again to \eqref{eq:new_tradeoff}. However, with the $\A$ structure given in Proposition~\ref{prop:Ab2} we can now select parameters that allow to exploit the trade-off in the regime $\TP< \MP/2+1$. The following proposition presents a structure for $\A$ which can be seen as combination of the structures from Propositions \ref{prop:Ab1} and \ref{prop:Ab2}, and which allows more freedom in the exploitation of the achievable trade-off in the regime $\TP\geq \MP/2+1$.

\begin{prop}\label{prop:Ab3}
Let $\A$ be given by \eqref{eq:A_kron}, with $\tA_\mathrm{T}=\mathbf{I}_{\TP}$, and $\tA_\mathrm{B}$ constructed as
\begin{equation}\label{eq:Ab3}
    \tA_\mathrm{B} =  \left[\begin{matrix}
         \boldsymbol{1}_{\Mdiff \times 1} & \left[\vec{1}_{Q_2 \times 1}\otimes\mathbf{I}_{J}\right]_{1:\Mdiff,:} & \underbrace{\begin{matrix}
         \boldsymbol{I}_{\Mdiff} &
         \cdots &
         \boldsymbol{I}_{\Mdiff} \end{matrix}
           }_{Q_1}
    \end{matrix}\right],
\end{equation}
where $Q_1\geq 1$ and $J$ are defined in Proposition~\ref{prop:Ab1}, while $Q_2$ is now given by
\begin{equation}\label{eq:Q3}
    Q_2 = \left\lceil \frac{\Mdiff}{J} \right\rceil.
\end{equation}
A randomly chosen matrix $\H$ admits WAX decomposition with probability 1 for the given $\A$ if
\begin{equation}\label{eq:cond_Ab3}
L \geq \frac{K}{1+Q_1+\frac{1}{Q_2}}.
\end{equation}
{Moreover, $\W_1$ (defined in \eqref{eq:w}) can be fixed to an arbitrary $L\times L$ full-rank matrix without affecting the solvability of the WAX decomposition.}
\begin{proof}
See Appendix~C
\end{proof}
\end{prop}
Note that for $Q_1=0$, the previous structure degenerates to the case from Proposition~\ref{prop:Ab2}, where some elements from the first column of $\tA_\mrm{B}$ should be changed to fulfill the additional $\alpha_i$ requirements. Furthermore, for $J=0$ (i.e., $\Mdiff$ divides $\TP-1$) the previous structure leads directly to the one presented in Proposition~\ref{prop:Ab1}. 

As happened in the previous cases, we can still reach the achievable trade-off \eqref{eq:new_tradeoff} whenever we have parameters such that $Q_2$ in \eqref{eq:Q3} evaluates to an integer value without the need of the ceiling operator. However, we can also reach it if we have parameters such that $Q_1$ evaluates to an integer value without the floor operation, since this would lead to $J=0$ and $Q_2$ would tend to infinity, so we could remove it altogether. Thus, the structure from Proposition~\ref{prop:Ab3} has a looser requirement so as to reach the achievable trade-off in the regime $\TP\geq\MP/2+1$ as compared to structure from Proposition~\ref{prop:Ab1}, where $Q_1$ had to evaluate to an integer value without the floor operation. Thus, the $\A$ structure defined in Proposition~\ref{prop:Ab3} allows for a broader selection of parameters leading to the achievable trade-off \eqref{eq:new_tradeoff}, hence increasing the freedom in the exploitation of said trade-off.

\subsection{General construction of valid $\Atil$}
A natural generalization of the structure given in Proposition~\ref{prop:Ab3}, which already corresponds to a generalization of the structures from Propositions~\ref{prop:Ab1} and \ref{prop:Ab2}, consists of filling the dimensions of $\tA_\mrm{B}$ with full identity matrices, alternating horizontal and vertical allocation until all dimensions are exhausted. This method is presented in Algorithm~\ref{alg:A_gen_const}, where the first column of $\tA_\mrm{B}$ is given by $\alpha_i$ so that it can accommodate degenerated cases as the one in Proposition~\ref{prop:Ab2}. The following conjecture aims at generalizing the validity of the structures defined by Algorithm~\ref{alg:A_gen_const}.

\begin{figure}[t!]
\begin{algorithm}[H]
\caption{Generalized $\Atil_\mrm{B}$ for WAX decomposition.}
 \begin{spacing}{1.4}
 \begin{algorithmic}\label{alg:A_gen_const}
 \REQUIRE $\MP$, $\TP$
 \ENSURE  $\Atil_\mrm{B}$ \\
 \textit{Initialize:}\\
  \hspace{1em} $\left[\Atil_\text{B}\right]_{1,:}\!=\!\begin{bmatrix}\alpha_1,\dots,\alpha_\Mdiff \end{bmatrix}^T$\\%\!,$\, \alpha_i\neq \alpha_j \text{ for }i\neq j$\\
  \hspace{1.2em}$R_\text{row}\!=\!\TP\!-\!1$,\; $R_\text{col}\!=\!\Mdiff,$ \;$i=0,$\, $i_\mrm{row}=1,$ \;$i_\mrm{col}=2$
 \WHILE{$R_\text{col}>0$ \AND $R_\text{row}>0$}
 \STATE $i = i+1$
 \IF{$R_\text{col}>R_\text{row}$}
 \STATE $Q_i = \left\lfloor R_\text{col}/R_\text{row} \right\rfloor$
 \STATE $[\Atil_\text{B}]_{i_\mrm{row}:(i_\mrm{row}+Q_i R_\text{row}),i_\mrm{col}:(i_\mrm{col}+Q_i R_\text{row})} \!=\!\vec{1}_{1\times Q_i}\otimes \mathbf{I}_{R_\text{row}}$
 \STATE $R_\text{col}=R_\text{col}-Q_i\cdot R_\text{row}$
 \STATE $i_\text{col}=i_\text{col}+Q_i\cdot R_\text{row}$
 \ELSE
 \STATE $Q_i = \left\lfloor R_\text{row}/R_\text{col} \right\rfloor$
 \STATE $[\Atil_\text{B}]_{i_\mrm{row}:(i_\mrm{row}+Q_i R_\text{col}), i_\mrm{col}:(i_\mrm{col}+Q_i R_\text{col})} \!=\! \vec{1}_{Q_i\times 1}\otimes \mathbf{I}_{R_\text{col}}$
 \STATE $R_\text{row}=R_\text{row}-Q_i\cdot R_\text{col}$
 \STATE $i_\text{row}=i_\text{row}+Q_i\cdot R_\text{col}$
 \ENDIF
 \ENDWHILE
 \end{algorithmic}
 \end{spacing}
 \end{algorithm}
\end{figure}

\begin{con}\label{con:Ab4}
Let $\A$ be given by \eqref{eq:A_kron}, with $\tA_\mathrm{T}=\mathbf{I}_{\TP}$, and $\tA_\mathrm{B}$ constructed through Algorithm~\ref{alg:A_gen_const}. A randomly chosen matrix $\H$ admits WAX decomposition with probability 1 for the given $\A$ if
\begin{equation}\label{eq:cond_Ab4}
    L\geq \frac{K}{1+Q_\mrm{tot}},
\end{equation}
where, given $Q_i$ for $i=1,\dots,N_\text{Q}$, which are defined in Algorithm~\ref{alg:A_gen_const}, and $N_\text{Q}$, corresponding to the iteration $i$ where dimensions are exhausted, we have 
\begin{equation}\label{eq:Q_cont}
    Q_\mrm{tot}= Q_1+\frac{1}{Q_2+\frac{1}{\ddots+\frac{1}{Q_{N_\text{Q}}}}}.
\end{equation}
Furthermore, in the regime $\TP<\MP/2+1$, the first column of $\tA_\mathrm{B}$, given by $[\alpha_1, \dots, \alpha_\Mdiff]^\mathrm{T}$, should fulfill the same restrictions as in Proposition~\ref{prop:Ab2}.
\begin{skproof}
    {We first note that $N_\text{Q}$ is determined by $\TP$ and $\MP$ (as more thoroughly discussed later), leading to integers in the range ${N_\text{Q}\in\{1,\dots,\mathrm{min}(\TP-1,\Phi)\}}$}. Then, for every value of $N_\text{Q}$ an equation similar to \eqref{eq:New_Ab3_form} can be obtained, which should be proved solvable. However, after extensive work on the matter, a formal proof for general $N_\text{Q}$ has not been found. We have only been able to test this formula through thorough simulations without encountering a single exception to it. { One simulation procedure we employed to check the conjecture was to randomly define a large number of combinations of $K$, $\MP$, and $\TP$, and for each of these combinations construct an $\boldsymbol{A}$ matrix through Algorithm~\ref{alg:A_gen_const} (together with \eqref{eq:A_kron} and \eqref{eq:At_Id}) using different values for $L$. Then, considering \cite[Theorem~2]{wax_journal}, we tried to perform WAX decomposition of a randomly chosen $\H$ (e.g., an IID Gaussian matrix realization), which would either be possible (i.e., $\A$ is valid) or not (i.e., $\A$ may not be valid). The simulation results led to valid $\A$ matrices if and only if Conjecture~\ref{con:Ab4} was satisfied.}
\end{skproof}
\end{con}

\begin{figure}[h]
\vspace{-0.8em}
	\centering
	\includegraphics[scale=0.55]{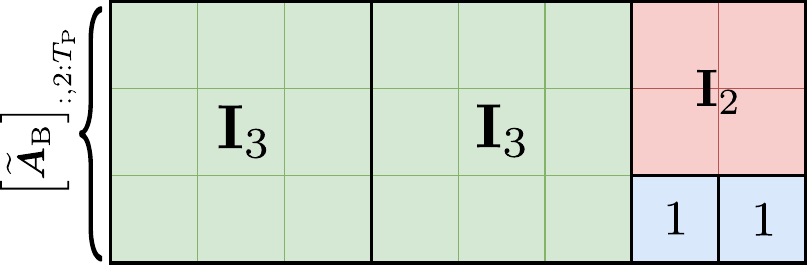}
\caption{Example of how Algorithm~\ref{alg:A_gen_const} constructs the last $\TP-1$ columns of $\widetilde{\A}_\mathrm{B}$ for $\MP=12$ and $\TP=9$.}
	\label{fig:euler}
\end{figure}

Figure~\ref{fig:euler} illustrates with an example how Algorithm~\ref{alg:A_gen_const} is used to define the $\TP-1$ last columns of $\widetilde{\A}_\mrm{B}$. We can immediately notice that its iterations are equivalent to the steps of the Euclidean algorithm for finding the greatest common divisor (GCD) between $\TP-1$ and $\Mdiff=\MP-\TP$. {In fact, we can see that $Q_\mrm{tot}$, given in \eqref{eq:Q_cont}, corresponds to a continued fraction expansion \cite{hardy75} of $(\TP-1)/\Mdiff$.} Hence, the value for $N_\text{Q}$, from Conjecture~\ref{con:Ab4}, is equal to the number of steps to calculate $\mathrm{GCD}(\TP-1,\Mdiff)$. {
Furthermore, since $\TP$ and $\MP$ are restricted to integers, $(\TP-1)/\Mdiff$ corresponds to a rational number, so its continued fraction expansion will always be finite \cite{hardy75}. Thus, we can substitute $Q_\mrm{tot}$ in \eqref{eq:cond_Ab4} by $(\TP-1)/\Mdiff$, which gives directly the achievable bound \eqref{eq:new_tradeoff}.} On the other hand, the number of 1s in the last $\TP-1$ columns of $\widetilde{\A}_\mrm{B}$, which gives its sparsity, corresponds to $$\left\Vert\left[\widetilde{\A}_\mrm{B}\right]_{:,2:\TP}\right\Vert^2_\mrm{F}=\Mdiff-\mathrm{GCD}(\TP-1,\Mdiff).$$

{The reader may also note here the direct relation between Conjecture~\ref{con:Ab4} and Propositions \ref{prop:Ab1}-\ref{prop:Ab3}. When $N_\mrm{Q}=1$, i.e., $\Mdiff$ divides $\TP-1$,  Conjecture~\ref{con:Ab4} directly corresponds to Proposition~\ref{prop:Ab1} (for this case we can choose $\alpha_i=1$). Furthermore, when $N_\mrm{Q}=2$, i.e., $J=\TP-1-Q_1\Mdiff$ divides $\Mdiff$, Conjecture~\ref{con:Ab4} leads to either Proposition~\ref{prop:Ab2} (in the $\TP<\MP/2+1$ regime) or Proposition~\ref{prop:Ab3} (in the $\TP\geq\MP/2+1$ regime, where we can choose $\alpha_i=1, \, \forall i$). Thus, although we lack a formal proof for Conjecture~\ref{con:Ab4}, we may use Algorithm~\ref{alg:A_gen_const} as a general strategy for constructing $\tA$ since it merges the previous results whenever they attain the achievable trade-off \eqref{eq:new_tradeoff}.}

{Another thing to remark is that, for centralized architectures, where we can identify $\TP=\MP$, our structures degenerate to the trivial $\tA=\mathbf{I}_{\MP}$, i.e., the combining module can be disregarded altogether. Moreover, for information-lossless fully-decentralized architectures (e.g., local-MF as in \cite{isit_2019}), where we can identify $\TP=1$ (for this case we have $L=K$), our structures degenerate to $\tA=\mathbf{1}_{\MP\times 1}$ (taking the case $\alpha_i=1$), i.e., the combining module would correspond to a  sum module that combines all the outputs from the decentralized filters. This remarks the relevance of the presented work as a generalization of architectures with an arbitrary level of decentralization, since it allows for a wide range of architectures from centralized to fully-decentralized, and where both extremes can be considered within the same framework.}

\section{Decentralized computation of $\W$}\label{section:decent}
The structures for $\A$ presented in the previous section are not only interesting for their sparsity and validity for WAX decomposition, but we can also use them to create decentralized schemes for computing $\W$. As previously discussed, decentralized here means that each panel would find the $\W_m$ to be applied, i.e., leading to information-lossless processing, by exchanging reduced data with the rest of the panels so that the full channel matrix $\H$ needs not be collected at any single point. Using the equivalent formulation of the WAX decomposition \eqref{eq:WAX_eq}, or \eqref{eq:WAX_IL} with \eqref{eq:A_kron}, we can find $\W_m$ matrices achieving information-lossless processing without having to compute $\X$. 

\begin{figure}[h]
	\centering
	\includegraphics[scale=0.476]{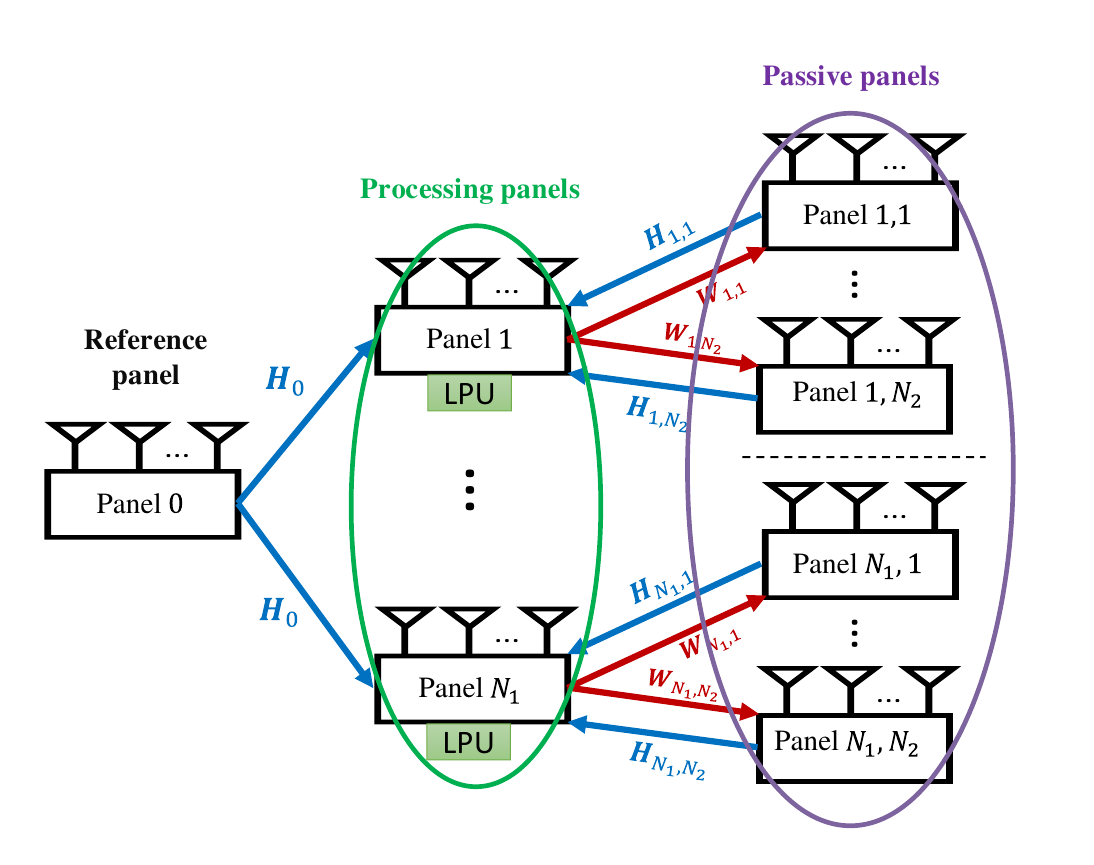}
\caption{
	\label{fig:sec:5} Architecture for decentralized computation of the $\W_m$ matrices for the $\tA$ given in Propositions~\ref{prop:Ab1}-\ref{prop:Ab3}. Blue arrows indicate sharing of local CSI, and red arrows indicate sharing of decentralized filters after computation.}
 \vspace{-0.2em}
\end{figure}

The $\A$ structures from Propositions \ref{prop:Ab1}, \ref{prop:Ab2}, and \ref{prop:Ab3}, allow the use of a tree-based scheme, such as the one illustrated in Fig.~\ref{fig:sec:5} where we have conveniently re-indexed the $\W_m$ and $\H_m$ matrices to make them general for all three cases. {Specifically, we identify now $\W_0$ with the original $\W_1$ from \eqref{eq:w}, which is the $\W_m$ that can be arbitrarily selected in Propositions~\ref{prop:Ab1}-\ref{prop:Ab3}.} The tree scheme consists of a reference panel, which is connected through a one-way link to $N_1$ processing panels, i.e., having a local processing unit (LPU), each of which communicates with a set of $N_2$ passive panels. {For simplicity, the reference panel makes use of the available freedom provided by Propositions~\ref{prop:Ab1}-\ref{prop:Ab3} by fixing $\W_0=\mathbf{I}_L$.} This way, $\W_0$ has no effect, so the reference panel only needs to share its $L\times K$ local channel matrix $\H_0$ with the $N_1$ processing panels.\footnote{{$\W_0$ can also be fixed to any other full-rank matrix. In that case, either all processing panels have previous knowledge of $\W_0$ for their computations, or the reference panel should share the $L\times K$ matrix resulting from the multiplication $\W_0^{-1}\H_0$ instead of $\H_0$ directly. Hence, any selection other than $\W_0=\mathbf{I}_L$ leads to higher computation complexity.}} Each group of $N_2$ passive panels would share their local channels to their corresponding processing panel, which would then use them to compute all the $\W_m$ matrices that have to be applied in its group (including itself). Lastly, the processing panels would send each $\W_m$ to the corresponding passive panels in their group so that they can apply them.

In order to understand why the $\A$ structures from Propositions~\ref{prop:Ab1}-\ref{prop:Ab3} can make use of the decentralized scheme from Fig.~\ref{fig:sec:5}, we will refer to the proofs of said propositions. For Proposition~\ref{prop:Ab1}, we can see that the equivalent formulation of the WAX decomposition can be solved by solving a set of independent equations of the form \eqref{eq:system_Ab1}, where the left-hand side (LHS), which is the only part shared in all equations, is associated to the reference panel ($\W_1$, here re-indexed to $\W_0$, which is later fixed in the proof), and the right-hand side (RHS) can be associated to a group of panels of which one would be the processing panel and the rest the passive panels. Each processing panel would only need the $\H_m$ matrices of the rest of the panels in the group, as well as the one from the reference panel, to be able to solve its equation, corresponding to one out of the $\Mdiff$ independent equations from \eqref{eq:system_Ab1}. For Proposition~\ref{prop:Ab2}, the reference panel determines the LHS of \eqref{eq:Ab2_problem}. On the other hand, \eqref{eq:Ab2_problem} can be divided into a set of independent equations of the form \eqref{eq:New_Ab2_form}, only sharing $\H_0$ (or $\widetilde{\H}_0$ in the proof), and each of which can be solved at one processing panel by accumulating the involved $\H_m$ matrices. The same is true for Proposition~\ref{prop:Ab3} where instead of \eqref{eq:New_Ab2_form} we would have \eqref{eq:New_Ab3_form} solved at each processing panel.

Table~\ref{tab:param_dec} gives the resulting parameters $N_1$ and $N_2$ of the decentralized scheme in Fig.\ref{fig:sec:5} for the different structures from Propositions~\ref{prop:Ab1}-\ref{prop:Ab3}. Said parameters are related to the number of independent equations and the number of involved passive panels in each equation, respectively, as explained before. In the case of Propositions \ref{prop:Ab2} and \ref{prop:Ab3}, we are assuming that $Q_2$ evaluates to an integer without the need of the ceiling operator; otherwise, the last group of panels would have a number of passive panels smaller than $N_2$ due to the cropping of the corresponding equation. Note that, in all cases, several independent equations can be solved at a single processing panel by gathering the corresponding $\H_m$ matrices at said panel. Thus, the values of $N_1$ from Table~\ref{tab:param_dec} could be trivially reduced by a corresponding increase in $N_2$.

{To conclude this section, we have shown that, not only can we define architectures with an arbitrary level of decentralization in the data phase (i.e., by employing the framework from Fig.~\ref{fig:gen_arch}), but, during the training phase, and if $\A$ is suitably selected, these architectures can be used for computing in a decentralized manner the decentralized processing to be applied (i.e., by considering schemes like the one in Fig.~\ref{fig:sec:5}).}
\begin{table}[]
    \renewcommand{\arraystretch}{1.2}
    \centering
    \begin{tabular}{|c||c|c|}
    \hline
         $\A$ structure & $N_1$ & $N_2$ \\
         \hline \hline
         Proposition~\ref{prop:Ab1} & $\Mdiff$ & $Q_1$\\
         \hline
         Proposition~\ref{prop:Ab2} & $\TP-1$ & $Q_2$ (from \eqref{eq:Q2})\\
         \hline
         Proposition~\ref{prop:Ab3} & $J$ & $Q_2$ (from \eqref{eq:Q3})\\
         \hline
    \end{tabular}
    \caption{Decentralized scheme parameters.}
    \label{tab:param_dec}
    \vspace{-0.8em}
\end{table}

{
\section{Numerical results and examples}}\label{section:ex_dis}
In Section~\ref{section:validA}, we presented some constructions for $\A$ that were proved to be valid for WAX decomposition. The current section aims at providing some discussion, as well as useful examples, to further understand the differences of said constructions, and the circumstances under which they reach the achievable trade-off \eqref{eq:new_tradeoff}.

\begin{figure}[h]
\vspace{-0.5em}
	\centering
	\includegraphics[scale=0.44]{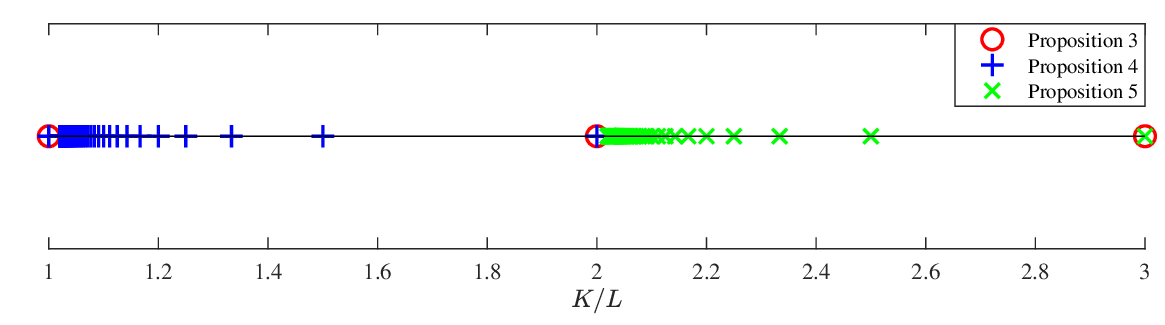}
\caption{Values of $K/L$ achieving \eqref{eq:new_tradeoff} in the interval $[1,3]$ with the $\A$ structures from the different propositions.}
\vspace{-0.5em}
\label{fig:K_L}
\end{figure}

In Section~\ref{section:validA}, we discussed  the requirements for Propositions~\ref{prop:Ab1}-\ref{prop:Ab3} to achieve \eqref{eq:new_tradeoff}, namely that either $Q_1$ or $Q_2$ should evaluate to an integer without requiring the floor/ceiling operator, respectively. Instead of obtaining $Q_1$ or $Q_2$ from $\TP$ and $\MP$, we can also take them as arbitrary integers, and substitute the resulting $K/L$ in \eqref{eq:cond_Ab1}, \eqref{eq:cond_Ab2}, and \eqref{eq:cond_Ab3} (after restricting the inequality for the case of equality) to get the ratios $K/L$ that achieve \eqref{eq:new_tradeoff} for the structures in Propositions \ref{prop:Ab1}, \ref{prop:Ab2}, and \ref{prop:Ab3}, respectively. The reason is that we can always find a combination of integers $\MP$ and $\TP$ leading to the corresponding $Q_1$ or $Q_2$ without the need of the respective floor/ceiling operators. {An alternative interpretation of the presented structures is that they are directly defined by a (truncated) continued fraction expansion of the ratio $K/L$, corresponding to \eqref{eq:Q_cont}. The structure from Algorithm~\ref{alg:A_gen_const} considers the full continued fraction expansion of $K/L$, the structure from Proposition~\ref{prop:Ab1} is given by a fraction expansion of $K/L$ truncated to a single term ($Q_1$), and the structure from Proposition~\ref{prop:Ab3} (which degenerates to the one from Proposition~\ref{prop:Ab2} for $Q_1=0$) is given by a fraction expansion of $K/L$ truncated to two terms ($Q_1$ and $Q_2$).} Fig.~\ref{fig:K_L} shows the possible $K/L$ ratios achieving \eqref{eq:new_tradeoff} for the $\A$ structures from Propositions~\ref{prop:Ab1}-\ref{prop:Ab3}. Proposition~\ref{prop:Ab2} is the only one having values in the interval $(1,2)$, associated to the regime $\TP< \MP/2+1$, as previously mentioned. However, the structure from Proposition~\ref{prop:Ab3} would also reach the points from Proposition~\ref{prop:Ab2} by  selecting the first column of $\widetilde{\A}_\mrm{B}$ as in \eqref{eq:Ab2} with $\alpha_i\neq \alpha_j$ for $i\neq j$.  Note that, for Proposition~\ref{prop:Ab3}, the values in the interval [2,3] can be shifted to any other interval [i,i+1] with $i\geq 2$ (which corresponds to increasing $Q_1$). Any other value for $K/L$ can be obtained by using Algorithm~\ref{alg:A_gen_const}, since any positive rational number can be decomposed into a continued fraction of the form \eqref{eq:Q_cont} \cite{hardy75}, while $K/L-1$ is inherently restricted to positive rational numbers since $K$ and $L$ are restricted to integers, and we have $L\leq K$. 

{
As previously discussed, the results presented in this paper lead to conditions that exhibit direct connection to the ratio $K/L$. However, the main value of the original condition \eqref{eq:cond_wax}, as presented in \cite{wax_journal}, and which has now been shown to be fundamental due to Theorem~\ref{thm:wax_conv}, is to give a trade-off between $L$, the number of multiplications per antenna, and $T$, the required connections to a CPU. It is thus of special interest to outline the explicit relation between $T$ and $L$ in the conditions obtained in this work: \eqref{eq:cond_Ab1}, \eqref{eq:cond_Ab2} and \eqref{eq:cond_Ab3}, as well as \eqref{eq:new_tradeoff}, which ultimately governs all the previous conditions (apart from being attained by the structure from Conjecture~\ref{con:Ab4}). On the other hand, the achievable trade-off from \eqref{eq:new_tradeoff} can be straightforwardly translated into a condition between $L$ and $T$ (instead of $\TP$) if we multiply both the numerator and denominator of the RHS by $L$. However, there may be points of this trade-off not attainable by the proposed structures, as we will illustrate next. %For the cases where Propositions~\ref{prop:Ab1}-\ref{prop:Ab3} do not attain \eqref{eq:new_tradeoff}, given $L$ and $K$ we can obtain the corresponding $T$ by fixing the $K/L$ to the closest cropped fraction expansion, i.e., by properly selecting $Q_1$ and/or $Q_2$, and obtaining the underlying $\TP$ as being the one fulfilling \eqref{eq:new_tradeoff} after substituting $K/L$ by its cropped fraction expansion.

In Fig.~\ref{fig:trade-off_new} we compare the trade-off between $T$ and $L$ considering the different strategies for constructing $\A$. The dashed blue line corresponds to the trade-off defined in \eqref{eq:cond_wax}, which is proved to be attained by randomly chosen $\A$ \cite[Theorem~1]{wax_journal}, while Theorem~\ref{thm:wax_conv} shows that it is also the optimum trade-off. The red line corresponds to the achievable trade-off \eqref{eq:new_tradeoff}, which can be attained by all the structures presented in this work under favorable parameter combinations. We can see that there is a minor gap between the achievable trade-off and the optimum one, which is mainly noticeable as $L$ grows. This gap can be explained by the exhaustion of degrees of freedom when fixing one $L\times L$ matrix, which clearly grows with $L$. The rest of the points correspond to the achievable points that can be exploited in practice through the proposed structures. {We have used \cite[Theorem~2]{wax_journal}, i.e., by performing WAX decomposition of a randomly chosen $\H$, to check that the proposed $\A$ structures are valid at these points, thus confirming the theoretical claims from Propositions~\ref{prop:Ab1}-\ref{prop:Ab3}, as well as Conjecture~\ref{con:Ab4}.} The purple triangles correspond to the achievable points with randomly chosen $\A$ after considering the integer restriction of the variables $T$, $L$, and $\MP$. Hence, these points correspond to the fundamental limits of multi-antenna architectures with an arbitrary level of decentralization, while the main motivation of the current work is to get as close as possible to these points with structured sparse constructions for $\A$. {The red circles correspond to the $\A$ structures defined through Proposition~\ref{prop:Ab1}, which was already included in the conference version \cite{icc2022}. The remaining points are novel contributions achieved by Propositions~\ref{prop:Ab2} and \ref{prop:Ab3}, as well as by Conjecture~\ref{con:Ab4}. As we can see, the achievable points when constructing $\A$ as in Conjecture~\ref{con:Ab4} have minor gap (if any) with respect to the achievable trade-off due to the integer nature of $\TP$ from the limitation $\TP=T/L$.} As for the other constructions, we see that the achievable points for the constructions from Propositions~\ref{prop:Ab2} and \ref{prop:Ab3} get fairly close to the points achieved by Conjecture~\ref{con:Ab4}, while the main difference is that they exploit different regimes of the trade-off. {We again note that the achievable points from Proposition~\ref{prop:Ab1} only allowed for reductions in the regime $\TP\geq \MP/2+1$, which justifies the poor performance in the right part of the plots. These results remark the importance of the novel structures presented in this work for better exploitation of the trade-off between level of decentralization and decentralized complexity.}

\begin{figure*}[h]
     \centering
     \begin{subfigure}[b]{0.5\textwidth}
         \centering
         \includegraphics[scale=0.54]{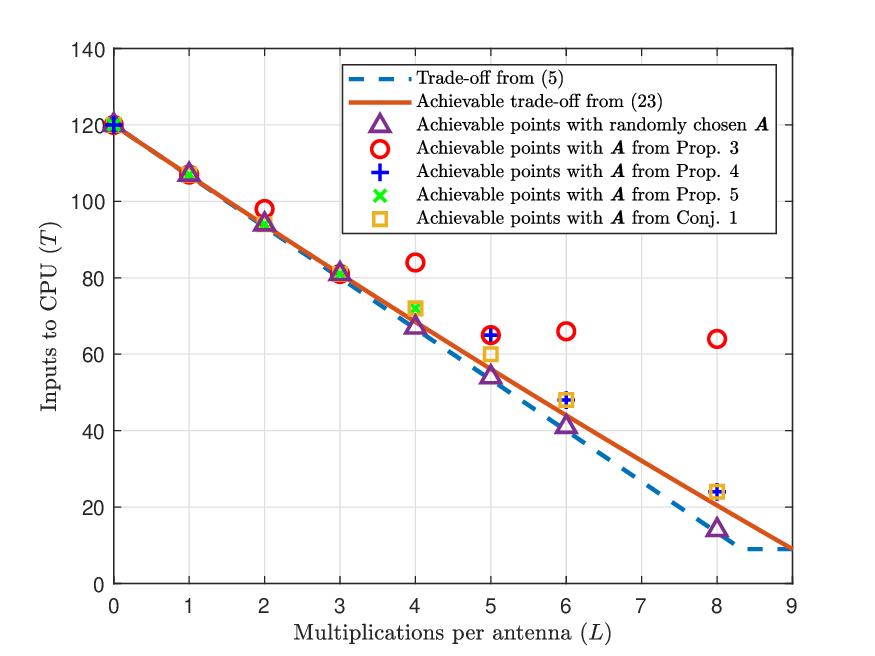}
     \end{subfigure}
     \hfill
     \begin{subfigure}[b]{0.49\textwidth}
         \centering
         \includegraphics[scale=0.54]{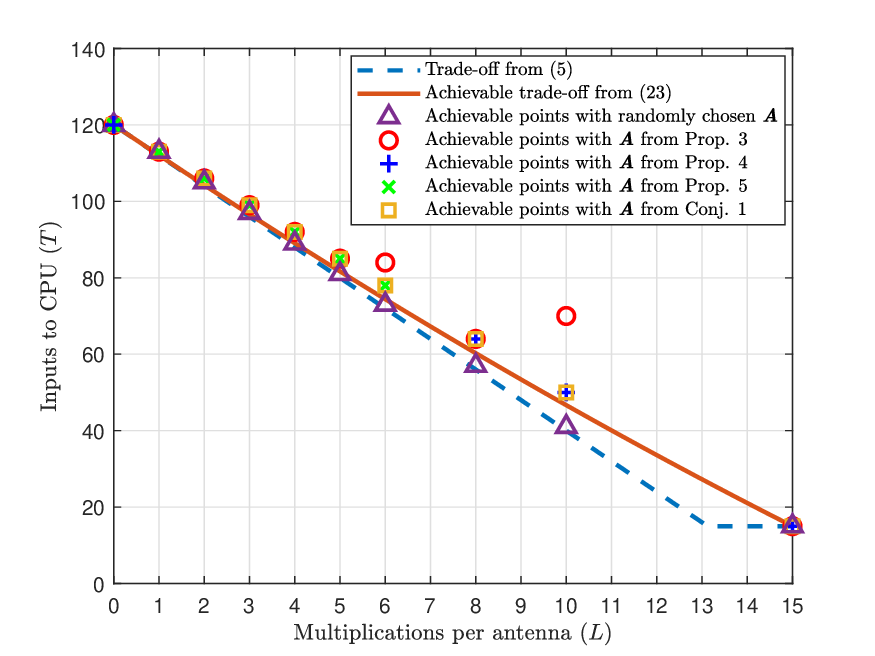}
     \end{subfigure}
     \caption{Comparison of fundamental trade-off \eqref{eq:cond_wax} with achievable trade-off \eqref{eq:new_tradeoff} and achievable points with the presented structures. We assume $M=120$, $K=9$ (left) and $K=15$ (right).}
        \label{fig:trade-off_new}
        \vspace{-0.8em}
\end{figure*}
}

The following example illustrates the differences between the presented strategies for constructing $\A$. By focusing on the regime $\TP\geq\MP/2+1$, we intentionally skip Proposition~\ref{prop:Ab2} due to its correspondence with Proposition~\ref{prop:Ab3} for $Q_1=0$.
\begin{ex}
Let $\TP=6$, $\MP=9$, and $K=40$. We then have $\Phi=3$. Let $\A_1$, $\A_2$, and $\A_3$, be $\A$ matrices constructed using Propositions~\ref{prop:Ab1}, \ref{prop:Ab3}, and Algorithm~\ref{alg:A_gen_const}, respectively. Such matrices are found in \eqref{ex}, where every pair of $\mathbf{I}_L$ matrices in each row block of $\A$ can be seen as a sum module which combines the outputs from the two respective panels.\footnote{We have horizontally flipped $[\widetilde{\A}_\mrm{B}]_{:,2:\TP}$ from Algorithm~\ref{alg:A_gen_const}, which is possible considering Propositions \ref{prop:inv_trans_A} and \ref{prop:perm_A},  to remark its similarity with the other constructions.} Note that the structure from Proposition~\ref{prop:Ab2} does not apply here since we are not in the regime $\TP<\MP/2+1$. Using Propositions~\ref{prop:Ab1}, \ref{prop:Ab3}, and Conjecture~\ref{con:Ab4}, we can find the possible values for $L$ to be
\begin{equation*}
    \begin{aligned}
    L_1\geq 20, \;\;\;\;\;\;   L_2\geq 16,  \;\;\;\;\;\;  L_3\geq 15,
    \end{aligned}
\end{equation*}
respectively. On the other hand, \eqref{eq:new_tradeoff} leads to the restriction $L\geq 15$. Thus, in this case Algorithm~\ref{alg:A_gen_const} gives the only structure reaching the achievable trade-off. However, the structure from Proposition~\ref{prop:Ab3} gets considerably closer to it than the one from Proposition~\ref{prop:Ab1}.
\end{ex}

We next show a practical example of how to use the theory developed in this work for designing a BS with constrained decentralized processing complexity.
\begin{ex}
Let us have a massive MIMO BS with $M=64$ antennas serving $K=10$ users. If we choose an arbitrary number of multiplications per antenna $L$, we would like to see which methods can be used for constructing a combining module $\A$, and what would be their resulting minimum number of inputs to the CPU. Recall that $L$ should be an integer dividing $M$ to be able to group the antennas into $\MP=M/L$ panels. The number of inputs to the CPU is given by $T=\TP L$, where we can find the minimum achievable integer $\TP$ from \eqref{eq:new_tradeoff} by
\begin{equation}\label{eq:min_Tp}
    T_{\mrm{P},\min} = \left\lceil \MP-\frac{M-L}{K} \right\rceil,
\end{equation}
which, using Conjecture~\ref{con:Ab4}, will always be possible by constructing $\A$ through Algorithm~\ref{alg:A_gen_const}.
\begin{itemize}
    \item Let us have $L=2$, which gives $\MP=32$ and the achievable trade-off $\TP\geq 25.8$ leading to $T_{\mrm{P},\min}=26$. If we choose to construct $\A$ through Proposition~\ref{prop:Ab1}, we start by using $\TP=T_{\mrm{P},\min}$ to get $\Mdiff=6$ from \eqref{def_of_Mdiff}. This gives $Q_1=4$ by \eqref{def_of_Q1}, which leads to the restriction $L\geq 2$. So we get the desired $L$ without the need to increase $\TP$. If we use Proposition~\ref{prop:Ab3} this restriction transforms to $L\geq 1.94$. Thus, for this case, due to the integer restrictions, there is no difference in terms of inputs to the CPU of defining $\A$ from Propositions \ref{prop:Ab1}, \ref{prop:Ab3}, or Algorithm~\ref{alg:A_gen_const},  since all give $T=52$. We would recommend Proposition \ref{prop:Ab1} for its simplicity and greater sparsity.
    \item Let us have $L=4$, which gives $\MP=16$ leading to $T_{\mrm{P},\min}=10$. If we want to construct $\A$ through Proposition~\ref{prop:Ab1}, we proceed as before using first $\TP=T_{\mrm{P},\min}$ to get $Q_1=1$, which leads to $L\geq 5$. In this case the desired $L$ is not possible, so we increase $\TP=T_{\mrm{P},\min}+1=25$, and calculate again $Q_1=2$, which leads to $L\geq 3.33$. This means that in order to use $\A$ from Proposition~\ref{prop:Ab1} we need $T=100$ inputs to the CPU. Instead, if we try to construct $\A$ from Proposition~\ref{prop:Ab3}, we start again with $\TP=T_{\mrm{P},\min}$, leading to $Q_1=2$ and $Q_2=2$. This gives the restriction $L\geq 4$, which is already fulfilled by the desired one. Thus, using Proposition~\ref{prop:Ab3} to define $\A$ would require $T=96$ inputs to the CPU, i.e., there is no loss with respect to the achievable gain, which can also be reached by defining $\A$ through Algorithm~\ref{alg:A_gen_const}.
\end{itemize}
\end{ex}

\begin{figure*}[b]
\vspace{-0.5em}
\hrulefill
\begin{equation} \label{ex}
\begin{matrix} \scalebox{0.8}{$\A_{\mrm{T}}$}\left\{ \begin{matrix} \\ \\ \\ \\ \\ \\ \end{matrix} \right. \\ \\ \\ \\ \end{matrix}\hspace{-2.4em}
\begin{matrix} \\ \\ \\ \\ \\ \\ \scalebox{0.8}{$\A_{\mrm{B}}$}\left\{ \begin{matrix} \\ \\ \\ \end{matrix} \right. \end{matrix}\hspace{-3.3em}\A_1 =\;\begin{bmatrix}
\mathbf{I}_L & \bs{0} & \bs{0} & \bs{0} & \bs{0} & \bs{0}\\
\bs{0} &\mathbf{I}_L & \bs{0} & \bs{0} & \bs{0} & \bs{0}\\
\bs{0} & \bs{0} & \mathbf{I}_L & \bs{0} & \bs{0} & \bs{0}\\
\bs{0} & \bs{0} & \bs{0} & \mathbf{I}_L & \bs{0} & \bs{0}\\
\bs{0} & \bs{0} & \bs{0} & \bs{0} & \mathbf{I}_L & \bs{0}\\
\bs{0}& \bs{0} & \bs{0} & \bs{0} & \bs{0} & \mathbf{I}_L\\
\mathbf{I}_L & \bs{0} & \bs{0} & \mathbf{I}_L & \bs{0} & \bs{0}\\
\mathbf{I}_L & \bs{0} & \bs{0} & \bs{0} & \mathbf{I}_L & \bs{0}\\
\mathbf{I}_L & \bs{0} & \bs{0} & \bs{0} & \bs{0} & \mathbf{I}_L
\end{bmatrix},\;\;
\A_2 =\begin{bmatrix}
\mathbf{I}_L & \bs{0} & \bs{0} & \bs{0} & \bs{0} & \bs{0}\\
\bs{0} &\mathbf{I}_L & \bs{0} & \bs{0} & \bs{0} & \bs{0}\\
\bs{0} & \bs{0} & \mathbf{I}_L & \bs{0} & \bs{0} & \bs{0}\\
\bs{0} & \bs{0} & \bs{0} & \mathbf{I}_L & \bs{0} & \bs{0}\\
\bs{0} & \bs{0} & \bs{0} & \bs{0} & \mathbf{I}_L & \bs{0}\\
\bs{0}& \bs{0} & \bs{0} & \bs{0} & \bs{0} & \mathbf{I}_L\\
\mathbf{I}_L & \mathbf{I}_L & \bs{0} & \mathbf{I}_L & \bs{0} & \bs{0}\\
\mathbf{I}_L & \bs{0} & \mathbf{I}_L & \bs{0} & \mathbf{I}_L & \bs{0}\\
\mathbf{I}_L & \mathbf{I}_L & \bs{0} & \bs{0} & \bs{0} & \mathbf{I}_L
\end{bmatrix},\;\;
\A_3 =\begin{bmatrix}
\mathbf{I}_L & \bs{0} & \bs{0} & \bs{0} & \bs{0} & \bs{0}\\
\bs{0} &\mathbf{I}_L & \bs{0} & \bs{0} & \bs{0} & \bs{0}\\
\bs{0} & \bs{0} & \mathbf{I}_L & \bs{0} & \bs{0} & \bs{0}\\
\bs{0} & \bs{0} & \bs{0} & \mathbf{I}_L & \bs{0} & \bs{0}\\
\bs{0} & \bs{0} & \bs{0} & \bs{0} & \mathbf{I}_L & \bs{0}\\
\bs{0}& \bs{0} & \bs{0} & \bs{0} & \bs{0} & \mathbf{I}_L\\
\mathbf{I}_L & \mathbf{I}_L & \bs{0} & \mathbf{I}_L & \bs{0} & \bs{0}\\
\mathbf{I}_L & \bs{0} & \mathbf{I}_L & \bs{0} & \mathbf{I}_L & \bs{0}\\
\mathbf{I}_L & \mathbf{I}_L & \mathbf{I}_L & \bs{0} & \bs{0} & \mathbf{I}_L
\end{bmatrix}
\end{equation}
\end{figure*}

\section{Conclusions}\label{section:conc}
We have continued with the work on WAX decomposition by filling some gaps from \cite{wax_journal}. {We have proved that the trade-off given in \cite{wax_journal} is fundamental in the sense that no decentralized system falling withing our general framework can perform beyond it.} We have defined an equivalent formulation of the WAX decomposition without the need of considering the CPU processing matrix $\X$. We have used said equivalent formulation to prove some properties that allow to transform the combining matrix $\A$ while maintaining its validity. We have also proved the validity of 3 structures for $\A$ which lead to an achievable version of the trade-off in \cite{wax_journal} under different system parameter settings. An ad hoc method for constructing $\A$ such that the achievable trade-off is reached for any system parameter setting is also presented. We have defined a decentralized scheme for obtaining the information-lossless decentralized filters $\W_m$ to be applied at different panels without the need to aggregate their CSIs.

Future work can include jointly considering the sparse combining modules $\A$ in scenarios where the channel matrix $\H$ is also sparse or has rank-deficiencies. More clever decentralized schemes, e.g., those which could be also employed with $\A$ matrices constructed through the general ad hoc method from Algorithm~\ref{alg:A_gen_const}, could also be explored. It would also be desirable to come up with a formal proof for the validity of the $\A$ matrices constructed through the ad hoc method.
{
\section*{Appendix A: Proof of Theorem~1}
The necessary condition $T\geq K$, in \eqref{eq:cond_wax} stated as $T>(K-1)$ due to the integer nature of $T$, comes trivially by the fact that $\mathrm{rank}(\W \A \X)\leq T$ and $\mathrm{rank}(\H)=K$ with probability 1 for randomly chosen $\H$. Let us thus assume $T\geq K$. If we invoke \cite[Lemma~3]{wax_journal}, we can conclude that a randomly chosen $\H$ admits WAX decomposition if and only if we can find full-rank $\W$ solving the linear system
    \begin{equation*}
        \A \X = \W^{-1} \H.
    \end{equation*}   
    The previous expression can be vectorized as in \cite{wax_journal} giving
    \begin{equation}\label{eq:vec_WAX}
        \left[\mathbf{I}_K\otimes \A \;\;\; -(\H^\mathrm{T}\otimes \mathbf{I}_M)\widetilde{\mathbf{I}}_{\W} \right]\begin{bmatrix}
            \mathrm{vec} (\X)\\ \mathrm{vec} (\W_1) \\ \vdots \\ \mathrm{vec} (\W_{\MP})
        \end{bmatrix}=\boldsymbol{0}_{MK\times 1},
    \end{equation}
    where $\widetilde{\mathbf{I}}_{\W}$ corresponds to an $M^2\times ML$ matrix having $\mathbf{I}_L$ blocks separated by blocks of zeros so as to disregard the zeros in $\mathrm{vec}(\W)$. The rank of the block $\mathbf{I}_K\otimes \A$, which multiplies $\mathrm{vec}(\X)$, is given by
    \begin{equation*}
        \mathrm{rank}(\mathbf{I}_K\otimes \A)=K R_{\A},
    \end{equation*}
    where $R_{\A}=\mathrm{rank}(\A)$. If $\mathrm{vec}(\X)$ is in the null-space of ${\mathbf{I}_K\otimes \A}$, then $[\mathrm{vec}(\W_1)^T,\dots,\mathrm{vec}(\W_{\MP})^T]^T$ should be in the null-space of $-(\H^\mathrm{T}\otimes \mathbf{I}_M)\widetilde{\mathbf{I}}_{\W}$ (full-rank with probability 1), leading to a more restrictive condition than \eqref{eq:cond_wax}, $K<L$. Thus, we can remove the subspace of $\mathrm{vec}(\X)$ that falls in the null-space of $\mathbf{I}_K\otimes \A$, which means that can rewrite \eqref{eq:vec_WAX} as
    \begin{equation}\label{eq:wax_vec2}
        \left[\boldsymbol{C} \;\; -(\H^\mathrm{T}\otimes \mathbf{I}_M)\widetilde{\mathbf{I}}_{\W} \right]\begin{bmatrix}
            \tilde{\boldsymbol{x}}\\ \mathrm{vec} (\W_1) \\ \vdots \\ \mathrm{vec} (\W_{\MP})
        \end{bmatrix}=\boldsymbol{0}_{MK\times 1},
    \end{equation}
    where $\boldsymbol{C}$ is now an $MK \times KR_{\A}$. Since $\H$ is a randomly chosen matrix, the block $-(\H^\mathrm{T}\otimes \mathbf{I}_M)\widetilde{\mathbf{I}}_{\W}$ adds full-rank to $\boldsymbol{C}$ with probability 1. Hence, the $MK \times (KR_{\A}+ML)$ matrix $\left[\boldsymbol{C} \;\; -(\H^\mathrm{T}\otimes \mathbf{I}_M)\widetilde{\mathbf{I}}_{\W} \right]$ is full-rank with probability 1, which means that it has non-empty null-space only if 
    \begin{equation}\label{eq:boundRA}
        MK <  KR_{\A}+ML.
    \end{equation}
    After simple manipulation of \eqref{eq:boundRA}, and noting that $R_{\A}\leq T$, where equality corresponds to $\A$ having rank T, we reach the necessary condition \eqref{eq:cond_wax} .
    }
\section*{Appendix B: Proof of Proposition~\ref{prop:Ab2}}
Selecting $\tA_\mathrm{T}=\mathbf{I}_{\TP}$ and $\tA_\mathrm{B}$ as in \eqref{eq:Ab2} leads to
\begin{equation}\label{eq:Btil2}
\Btil = \begin{bmatrix}\begin{matrix}\alpha_1 \vec{1}_{(\TP-1) \times 1} & \mathbf{I}_{\TP-1}\\
    \vdots & \vdots\\
    \alpha_{Q_2-1} \vec{1}_{(\TP-1)\times 1} & \mathbf{I}_{\TP-1}\\
    \alpha_{Q_2} \vec{1}_{\Pi\times 1} & \left[\mathbf{I}_{\TP-1}\right]_{1:\Pi,:}
    \end{matrix}&\scalebox{1.5}{$ -\mathbf{I}_\Mdiff $}\end{bmatrix}^\mathrm{T}.
\end{equation}
%$$
%\Btil = \begin{bmatrix} \begin{matrix}\alpha_1\otimes %\vec{1}_{1\times(\TP-1)} & \cdots &  \alpha_{Q_2-1}\otimes %\vec{1}_{1\times (\TP-1)} & \alpha_{Q_2}\otimes \vec{1}_{1\times \Pi}\\
    %\mathbf{I}_{\TP-1} & \cdots & \mathbf{I}_{\TP-1} & %\left[\mathbf{I}_{\TP-1}\right]_{:,1:\Pi}
%    \end{matrix} \\
%    \scalebox{1.5}{$ -\mathbf{I}_\Mdiff $}\end{bmatrix}.
%$$
If we use Corollary~\ref{cor:Akron}, we can substitute $\Btil$ in the equivalent formulation of the WAX decomposition, given in \eqref{eq:WAX_IL_face}. We then fix $\W_1^{-1}$ to some arbitrary full-rank matrix, for simplicity let us have $\W_1^{-1}=\mathbf{I}_L$ {(any other full rank-matrix can be absorbed by $\H_1$ or by the remaining $\W_m$'s)}, so \eqref{eq:WAX_IL_face} gives
\begin{equation}\label{eq:Ab2_problem}
\begin{aligned}
    -\begin{bmatrix}\alpha_1 & \cdots & \alpha_{Q_2}\end{bmatrix}  \otimes  \H_1 = \begin{bmatrix}\W_2^{-1} & \cdots & \W_{\MP}^{-1}
    \end{bmatrix}&\\
    \times \left( \begin{bmatrix}\begin{matrix} \mathbf{I}_{\TP-1} & \cdots &  \left[\mathbf{I}_{\TP-1}\right]_{1:\Pi,:}
    \end{matrix} \\
    \scalebox{1.3}{$ -\mathbf{I}_\Mdiff $}\end{bmatrix} \bullet \begin{bmatrix}
    \H_2\\
    \vdots \\
    \H_{\MP}
    \end{bmatrix}\right)&.
\end{aligned}
\end{equation}
We can notice that the face-splitting product $(.)\bullet(.)$ only substitutes in the left matrix each 1 at row $m$ by $\H_m$. Furthermore, \eqref{eq:Ab2_problem} corresponds to a series of $\TP-1$ independent equations of the form
\begin{equation}
\begin{aligned}\label{eq:New_Ab2_form}
   \begin{bmatrix}\alpha_1 & \cdots & \alpha_{Q_2}\end{bmatrix}  \otimes  \widehat{\H}_0 =  \begin{bmatrix}\widehat{\W}_1^{-1} &  \cdots  & \widehat{\W}_{Q_2+1}^{-1}
    \end{bmatrix}&\\
     \times \begin{bmatrix}
    \widehat{\H}_1 & \widehat{\H}_1 & \cdots & \widehat{\H}_1 \\
    \widehat{\H}_2 & \vec{0}_{L\times K} &  \cdots & \vec{0}_{L\times K}\\
    \vec{0}_{L\times K} & \widehat{\H}_3 & \ddots & \vdots \\
     \vdots & \ddots & \ddots & \vec{0}_{L\times K} \\
     \vec{0}_{L\times K}& \cdots & \vec{0}_{L\times K} & \widehat{\H}_{Q_2+1}
    \end{bmatrix}&,
\end{aligned}
\end{equation}
where $\widehat{\H_i}$ corresponds to a re-indexing of the respective $\H_m$, including a possible change of sign ($\widehat{\H}_0=-\H_1$ is the only matrix shared in all equations), so we can think of them as $L \times K$ blocks from a randomly chosen matrix. Note that we may also require the ability to solve a sub-problem of \eqref{eq:New_Ab2_form} with {$1$ less column block}, i.e., substituting $Q_2$ by $Q_2-1$. This is due to the possibly cropped block in \eqref{eq:Ab2} or {\eqref{eq:Ab2_problem}, $\left[\mathbf{I}_{\TP-1}\right]_{1:\Pi,:}$, which would lead to $(\TP-1-\Pi)$ equations having $Q_2-1$ instead of $Q_2$ column blocks in \eqref{eq:New_Ab2_form}. However, }this sub-problem is less restrictive than \eqref{eq:Ab2}, as we will see.

Let us multiply from the right both sides of \eqref{eq:New_Ab2_form} by the full rank matrix $\mathrm{diag}(\widehat{\vec{V}}_1, \dots, \widehat{\vec{V}}_1)$, where $\widehat{\vec{V}}_1$ is the $K \times K$ right unitary matrix from the singular value decomposition of $\widehat{\H}_1$. If we further use the fact that any full-rank block diagonal matrix being multiplied between the $\widehat{\W}_m^{-1}$'s and $\widehat{\H}_m$'s block matrices in the RHS of \eqref{eq:New_Ab2_form}, as well as any full-rank $L\times L$ matrix that multiplies from the left both sides of \eqref{eq:New_Ab2_form}, can be absorbed by the corresponding $\widehat{\W}_m^{-1}$ matrices, we reach
\begin{equation}
\begin{aligned}\label{eq:New_Ab2_form2}
    ([\alpha_1 \;\, \cdots \;\, \alpha_{Q_2}] \otimes [\mathbf{I}_L \;\; \widetilde{\H}_0]  ) =  \begin{bmatrix}\widetilde{\W}_1^{-1} & \! \cdots \! & \widetilde{\W}_{Q_2+1}^{-1}
    \end{bmatrix}&\\
    \scalebox{0.9}{$\times \begin{bmatrix}
    [\mathbf{I}_L \;\vec{0}_{L\times(K-L)}] & [\mathbf{I}_L \; \vec{0}_{L\times(K-L)}] & \cdots & [\mathbf{I}_L \; \vec{0}_{L\times(K-L)}] \\
    [\mathbf{I}_L \; \widetilde{\H}_2] & \vec{0}_{L\times K} &  \cdots & \vec{0}_{L\times K}\\
    \vec{0}_{L\times K} & [\mathbf{I}_L \; \widetilde{\H}_3] & \ddots & \vdots \\
     \vdots & \ddots & \ddots & \vec{0}_{L\times K} \\
     \vec{0}_{L\times K}& \cdots & \vec{0}_{L\times K} & [\mathbf{I}_L \; \widetilde{\H}_{Q_2+1}]
    \end{bmatrix}$}&,
\end{aligned}
\end{equation}
where $\widetilde{\H}_i$, $i\in \{0, 2, \dots, (Q_2+1)\}$, are now $L \times (K-L)$ blocks from a randomly chosen matrices.\footnote{Note that the multiplication by a common unitary matrix from the right to generate each $\widetilde{\H}_i$ can be seen as a common rotation to their original random unitary matrices, thus it does not affect the randomly chosen property} Equation \eqref{eq:New_Ab2_form2} corresponds to the system of equations
\begin{equation}\label{eq:set_eq_Ab2}
    \left\{\begin{array}{l}
    \widetilde{\W}_1^{-1}+\widetilde{\W}_{i+1}^{-1}=\alpha_{i}\mathbf{I}_L \\
    \widetilde{\W}_{i+1}^{-1}\widetilde{\H}_{i+1}=\alpha_{i}\widetilde{\H}_0
    \end{array}\right.,\;\;\; i=1,\dots, Q_2.
\end{equation}
{We can now isolate $\widetilde{\W}_{i+1}^{-1}=\alpha_i\mathbf{I}_L-\widetilde{\W}_1^{-1}$ from the first line of \eqref{eq:set_eq_Ab2}, and then substitute it in the second line to reach $\alpha_i\widetilde{\H}_{i+1}-\widetilde{\W}_1^{-1}\widetilde{\H}_{i+1}=\alpha_{i}\widetilde{\H}_0$. After reordering terms and merging the inequalities for $i=1,\dots, Q_2$, into matrix notation, we can write}
\begin{equation}\label{eq:W1_Ab2}
\widetilde{\W}_1^{-1}\widetilde{\vec{\mathcal{H}}} = \begin{bmatrix}\alpha_1(\widetilde{\H}_{2}-\widetilde{\H}_0) & \cdots & \alpha_{Q_2}(\widetilde{\H}_{Q_2+1}-\widetilde{\H}_0)\end{bmatrix},
\end{equation} 
where $\widetilde{\vec{\mathcal{H}}} = \begin{bmatrix}\widetilde{\H}_{2}& \cdots & \widetilde{\H}_{Q_2+1}\end{bmatrix}$ gives an $L\times Q_2(K-L)$ randomly chosen matrix, which is thus full-rank with probability 1. The block matrix on the RHS of \eqref{eq:W1_Ab2} is also a randomly chosen matrix, and thus full-rank with probability 1, as long as $\alpha_i \neq 0$ $\forall i$. We now note that \eqref{eq:W1_Ab2} corresponds to a linear equation solvable for $L\geq Q_2(K-L)$, which directly gives us the condition \eqref{eq:cond_Ab2}. 

It remains to prove that we have a full-rank solution for each $\widetilde{\W}_i^{-1}$ (then each corresponding $\W_m^{-1}$ would also be full-rank). Solving for $\widetilde{\W}_1^{-1}$ in \eqref{eq:set_eq_Ab2} gives the set of solutions
\begin{equation}\label{eq:W1_isol_Ab2}
%\begin{aligned}
\widetilde{\W}_1^{-1}\! =\! \begin{bmatrix}\alpha_1(\widetilde{\H}_{2}-\widetilde{\H}_0)  \cdots  \alpha_{Q_2}(\widetilde{\H}_{Q_2+1}-\widetilde{\H}_0)\end{bmatrix}\widetilde{\vec{\mathcal{H}}}^\dagger+\mathbf{N}_{\widetilde{\vec{\mathcal{H}}}},
%\end{aligned}
\end{equation}
where $\mathbf{N}_{\widetilde{\vec{\mathcal{H}}}}$ can be selected to be any $L\times L$ matrix in the left null-space of $\widetilde{\vec{\mathcal{H}}}$. Thus, $\mathrm{rank}(\mathbf{N}_{\widetilde{\vec{\mathcal{H}}}})\leq L-Q_2(K-L)$, and $\mathbf{N}_{\widetilde{\vec{\mathcal{H}}}}$ would vanish in case of equality in \eqref{eq:cond_Ab2} ($\widetilde{\vec{\mathcal{H}}}$ square). We can now note that the first term in the sum from the RHS of \eqref{eq:W1_isol_Ab2} has rank $Q_2(K-L)$ with probability 1. The reason is that it is the multiplication of an $L\times Q_2(K-L)$ matrix with a $Q_2(K-L)\times L$ matrix, so its rank cannot be above $Q_2(K-L)$, while, if we multiply $\widetilde{\vec{\mathcal{H}}}$ from the right, which cannot increase the rank, we get an $L\times Q_2(K-L)$ randomly chosen matrix (full-rank with probability 1). On the other hand, $\mathbf{N}_{\widetilde{\vec{\mathcal{H}}}}$ adds its rank to the other term of the sum, since they are in perpendicular spaces (left null-space and row-space are perpendicular). Therefore, by selecting any $\mathbf{N}_{\widetilde{\vec{\mathcal{H}}}}$ spanning the whole null-space of $\widetilde{\vec{\mathcal{H}}}$, i.e., having rank $L-Q_2(K-L)$, we  get a full-rank  $\widetilde{\W}_1^{-1}$.

We now show that full-rank solutions for $\widetilde{\W}_i^{-1}$, $i> 1$, are also available as long as $\alpha_i\neq \alpha_j$ for $i\neq i$. Substituting $\widetilde{\W}_1^{-1}$ from \eqref{eq:W1_Ab2} in the first equation of \eqref{eq:set_eq_Ab2} gives a solution for each $\widetilde{\W}_i^{-1}$ of the form
\begin{equation}
\begin{aligned}
    \widetilde{\W}_i^{-1} =& \alpha_i \mathbf{I}_L-\mathbf{N}_{\widetilde{\vec{\mathcal{H}}}}\\
    &\!-\!\begin{bmatrix}\alpha_1(\widetilde{\H}_{2}-\widetilde{\H}_0) & \cdots & \alpha_{Q_2}(\widetilde{\H}_{Q_2+1}-\widetilde{\H}_0)\end{bmatrix}\widetilde{\vec{\mathcal{H}}}^\dagger.
\end{aligned}
\end{equation}
Let us define $\tilde{\alpha}_j=\alpha_j-\alpha_i$. We then have
\begin{equation}
\begin{aligned}
    \widetilde{\W}_i^{-1} =& \alpha_i \left( \mathbf{I}_L-\widetilde{\vec{\mathcal{H}}}\widetilde{\vec{\mathcal{H}}}^\dagger\right)-\mathbf{N}_{\widetilde{\vec{\mathcal{H}}}} \\
    &-\begin{bmatrix}\tilde{\alpha_1}\widetilde{\H}_{2} & \cdots & \tilde{\alpha}_{Q_2}\widetilde{\H}_{Q_2+1}\end{bmatrix}\widetilde{\vec{\mathcal{H}}}^\dagger\\
    &+\begin{bmatrix}\alpha_1\widetilde{\H}_0 & \cdots & \alpha_{Q_2}\widetilde{\H}_0\end{bmatrix}\widetilde{\vec{\mathcal{H}}}^\dagger.
\end{aligned}
\end{equation}
However, it can be checked that
\begin{equation*}
    \widetilde{\vec{\mathcal{H}}}\widetilde{\vec{\mathcal{H}}}^{\dagger} = \vec{U}\mathrm{diag}\left(\mathbf{I}_{Q_2(K-L)}, \vec{0}_{L-Q_2(K-L)}\right) \vec{U}^{\mathrm{H}},
\end{equation*}
where $\vec{U}$ corresponds to the left unitary matrix from the singular value decomposition of $\widetilde{\vec{\mathcal{H}}}$.
Thus, we get
\begin{equation}\label{eq:Wi_Ab2}
\begin{aligned}
    \widetilde{\W}_i^{-1} = \alpha_i \vec{U} \mathrm{diag}\left(\vec{0}_{Q_2(K-L)}, \mathbf{I}_{L-Q_2(K-L)}\right) \vec{U}^\mathrm{H}-\mathbf{N}_{\widetilde{\vec{\mathcal{H}}}}& \\
    -\begin{bmatrix}\tilde{\alpha_1}\widetilde{\H}_{2} & \cdots & \tilde{\alpha}_{Q_2}\widetilde{\H}_{Q_2+1}\end{bmatrix}\widetilde{\vec{\mathcal{H}}}^\dagger&\\
    +\begin{bmatrix}\alpha_1\widetilde{\H}_0 & \cdots & \alpha_{Q_2}\widetilde{\H}_0\end{bmatrix}\widetilde{\vec{\mathcal{H}}}^\dagger&.
\end{aligned}
\end{equation}
We then note that the first two matrices on the RHS of \eqref{eq:Wi_Ab2} are both in the null-space of $\widetilde{\vec{\mathcal{H}}}$, and, since we have freedom in selecting $\mathbf{N}_{\widetilde{\vec{\mathcal{H}}}}$ as long as it spans the whole null-space, we can choose it so that the rank from the first matrix is not reduced after subtracting. Therefore, the first two matrices will always add rank $L-Q_2(K-L)$ to the last two, which lay in the row-space of $\widetilde{\vec{\mathcal{H}}}$. On the other hand, after multiplying $\widetilde{\vec{\mathcal{H}}}$ to the last two matrices in the RHS of \eqref{eq:Wi_Ab2} we get one matrix of rank $N_{\tilde{\alpha}} (K-L)$, with $N_{\tilde{\alpha}}$ the number of non-zero $\tilde{\alpha}_j$, and one matrix of rank $(K-L)$. Note that $N_{\tilde{\alpha}}\leq (Q_2-1)$ since $\tilde{\alpha}_i=0$ by definition. The sum of the latter two matrices would then be
$$\begin{bmatrix}\alpha_1\widetilde{\H}_0\!-\!\tilde{\alpha_1}\widetilde{\H}_{2} \!& \cdots &\! \alpha_i\widetilde{\H}_0 &\! \cdots &\! \alpha_{Q_2}\widetilde{\H}_0\!-\!\tilde{\alpha}_{Q_2}\widetilde{\H}_{Q_2+1}\end{bmatrix}.
$$
The rank is then\footnote{$\widetilde{\H}_0$ and $\widetilde{\H}_{i}$ do not destroy rank since they are blocks from a randomly chosen matrix, and they are full-rank with probability 1.} $(N_{\tilde{\alpha}}+1) (K-L)$ so, in order to have \eqref{eq:Wi_Ab2} full-rank, we need $N_{\tilde{\alpha}}=Q_2-1$ for each $i$, which means all $\tilde{\alpha_j}$ ($j\neq i$) should be non-zero. Considering $\widetilde{\W}_i^{-1}$ $\forall i$, this translates to having $\alpha_i\neq \alpha_j$ for $i\neq j$. Hence, Proposition~\ref{prop:Ab2} is proved.
%Let us define the singular value decomposition of each of the $L\times K$ blocks of the channel matrix as
%$$
%\H_m = \boldsymbol{U}_m \begin{bmatrix} \boldsymbol{S}_m & \boldsymbol{0}_{L \times (K-L)} \end{bmatrix} \boldsymbol{V}_m^\mathrm{H} , \; m=1,\dots,\MP
%$$
%where $\boldsymbol{U}_m$ and $\boldsymbol{V}_m$ are $L \times L$ and $K \times K$ unitary matrices, respectively, and $\boldsymbol{S}_m$ is an $L \times L$ diagonal matrix containing the singular values of $\H_m$ along its diagonal.

\section*{Appendix C: Proof of Proposition~\ref{prop:Ab3}}
Selecting $\tA_\mathrm{T}=\mathbf{I}_{\TP}$ and $\tA_\mathrm{B}$ as in \eqref{eq:Ab3} leads to
$$
\Btil = \begin{bmatrix} \boldsymbol{1}_{\Mdiff \times 1} & \left[\vec{1}_{Q_2 \times 1}\otimes\mathbf{I}_{J}\right]_{1:\Mdiff,:} &\underbrace{\mathbf{I}_\Mdiff \;\; \cdots \;\;  \mathbf{I}_\Mdiff }_{Q_1} \;\; -\mathbf{I}_\Mdiff \end{bmatrix}^\mathrm{T}.
$$
Note the similarity of the previous $\widetilde{\vec{B}}$ with \eqref{eq:Btil2}, which for $\alpha_i=1$ only the $Q_1$ extra $\mathbf{I}_\Mdiff$ would be added. Applying similar arguments as in the proof of Proposition~\ref{prop:Ab2}, {which include fixing $\W_1$ to an arbitrary full-rank matrix}, we can transform the equivalent formulation of the WAX decomposition \eqref{eq:WAX_IL_face} into a series of independent equations of the form
\begin{equation}\label{eq:New_Ab3_form}
\begin{aligned}
  \boldsymbol{1}_{1\times Q_2}  \otimes  \widehat{\H}_0 =  \begin{bmatrix}\widehat{\W}_1^{-1} \! & \! \cdots \! & \! \widehat{\W}_{(Q_1+1) Q_2+1}^{-1}&
    \end{bmatrix} \\
    \times \begin{bmatrix}
    \widehat{\H}_1 &  \cdots & \widehat{\H}_1 \\
    \widehat{\H}_2 &  &  \\
     & \ddots &  \\
     & & \widehat{\H}_{Q_2+1} \\
     & \vdots & \\
      \widehat{\H}_{Q_1 Q_2+2}& & \\
      & \ddots & \\
      & & \widehat{\H}_{(Q_1+1) Q_2+1}\\
    \end{bmatrix}&,
\end{aligned}
\end{equation}
where we have relaxed notation by removing the blocks of zeros. Again, only $\widehat{\H}_0$ is shared among the different independent equations. Let us have
$$
\widehat{\H}_m = \begin{bmatrix} \widehat{\H}_{m,\mathrm{sq}} & \widehat{\H}_{m,\mathrm{r}}
\end{bmatrix}, \; m=0, \dots, (Q_1+1)Q_2+1,
$$
where $\widehat{\H}_{m,\mathrm{sq}}$ and $\widehat{\H}_{m,\mathrm{r}}$  are $L\times L$ and $L\times (K-L)$ blocks from a randomly chosen matrix, respectively. We can then have $\widehat{\H}_{1,\mathrm{r}}=\boldsymbol{0}_{L, (K-L)}$ by absorbing the corresponding right unitary matrix in the rest of $\widehat{\H}_m$ as before. We then get the set of equations
\begin{equation}\label{eq:set_eq_Ab3}
    \left\{\begin{array}{l}
    \widehat{\W}_1^{-1}\widehat{\H}_{1,\mathrm{sq}}+\sum_{q=0}^{Q_1}\widehat{\W}_{i+1+q Q_2}^{-1}\widehat{\H}_{i+1+q Q_2,\mathrm{sq}}=\widehat{\H}_{0,\mathrm{sq}} \\
    \sum_{q=0}^{Q_1}\widehat{\W}_{i+1+q Q_2}^{-1}\widehat{\H}_{i+1+q Q_2,\mathrm{r}}=\widehat{\H}_{0,\mathrm{r}}
    \end{array}\right.,
\end{equation}
where $i=1,\dots, Q_2$. Let us isolate $\widetilde{\W}_{i+1+Q_1 Q_2}^{-1}$ in the first equation of \eqref{eq:set_eq_Ab3}
\begin{equation}
\begin{aligned}
\widehat{\W}_{i+1+Q_1 Q_2}^{-1} &= \bigg(\widehat{\H}_{0,\mathrm{sq}}-\widehat{\W}_1^{-1}\\
-&\sum_{q=0}^{Q_1-1}\widehat{\W}_{i+1+q Q_2}^{-1}\widehat{\H}_{i+1+q Q_2,\mathrm{sq}}\bigg)\widehat{\H}_{i+1+Q_1 Q_2,\mathrm{sq}}^{-1},
\end{aligned}
\end{equation}
which, assuming full-rank  $\widehat{\W}_{i+1+q Q_2}$ for $q<Q_1$, corresponds to a random combination of full-rank matrices, so it will lead to full-rank $\widehat{\W}_{i+1+Q_1 Q_2}^{-1}$ with probability 1. Substituting in the second equation from \eqref{eq:set_eq_Ab3}, absorbing some square randomly chosen matrices (full-rank with probability 1) in the corresponding $\widehat{\W}_i$, and renaming blocks, we get
\begin{equation}
\begin{aligned}
\widetilde{\W}_1^{-1}\widetilde{\H}_{1i}+\sum_{q=0}^{Q_1-1}\widetilde{\W}_{i+1+q Q_2}^{-1}\widetilde{\H}_{i+1+q Q_2}=\widetilde{\H}_{0}+\widetilde{\H}_{1i},
\end{aligned}
\end{equation}
where all $\widetilde{\H}_{m}$ (or $\widetilde{\H}_{mn}$) correspond again to blocks of size $L \times (K-L)$ from a randomly chosen since they come from sums and products of different blocks from a randomly chosen matrix. Multiplying both sides by $\widetilde{\vec{V}}_{1i}$, where $\widetilde{\vec{V}}_{1i}^\mathrm{H}$ corresponds to the right unitary matrix of $\widetilde{\H}_{1i}$, we reach
\begin{equation}\label{eq:New_Ab3_1st_red}
\begin{aligned}
  \scalebox{0.95}{$\begin{bmatrix}\!\widetilde{\H}_{01}+\widetilde{\H}_{11} \! & \! \cdots \! & \! \widetilde{\H}_{0Q_2}+\widetilde{\H}_{1Q_2}
    \end{bmatrix} \!=\!   \begin{bmatrix}\widetilde{\W}_1^{-1} \! & \! \cdots \! & \! \widetilde{\W}_{Q_1 Q_2\!+\!1}^{-1}
    \end{bmatrix}$}& \\
    \times  \begin{bmatrix}
    \widetilde{\H}_{11} &  \cdots & \widetilde{\H}_{1Q_2} \\
    \widetilde{\H}_2 &  &  \\
     & \ddots &  \\
      & & \widetilde{\H}_{Q_2+1}\\
       & \vdots &  \\
      \widetilde{\H}_{(Q_1-1) Q_2+2}& & \\
      & \ddots & \\
      & & \widetilde{\H}_{Q_1 Q_2+1}\\
    \end{bmatrix}&,
\end{aligned}
\end{equation}
where $\widetilde{\H}_{0i} = \widetilde{\H}_{0}\widetilde{\vec{V}}_{1i}$, and $\widetilde{\H}_{1i}=[\widetilde{\H}_{1i,\mathrm{sq}} \;\; \boldsymbol{0}_{L\times (K-2L)}]$. We then reach the following set of equations for $i=1,\dots, Q_2$
\begin{equation}\label{eq:set2_eq_Ab3}
    \left\{\begin{array}{l}
    \widetilde{\W}_{1}^{-1}\widetilde{\H}_{1i,\mathrm{sq}}\!+\!\sum_{q=0}^{Q_1\!-1}\!\widetilde{\W}_{i+1+q Q_2}^{-1}\widetilde{\H}_{i+1+q Q_2,\mathrm{sq}}\!=\!\widetilde{\H}_{0i,\mathrm{sq}} \\
    \sum_{q=0}^{Q_1-1}\widetilde{\W}_{i+1+q Q_2}^{-1}\widetilde{\H}_{i+1+q Q_2,\mathrm{r}}=\widetilde{\H}_{0i,\mathrm{r}}
    \end{array}\right.\!,
\end{equation}
where $\widetilde{\H}_{m}=[\widetilde{\H}_{m,\mathrm{sq}} \;\; \widetilde{\H}_{m,\mathrm{r}}]$, with $\widetilde{\H}_{m,\mathrm{sq}}$ being square blocks as before. Note that \eqref{eq:set2_eq_Ab3} is almost like \eqref{eq:set_eq_Ab3}, but the dimensions have been reduced, as well as the number of sum elements, and we have now different $\widetilde{\H}_{1i,\mathrm{sq}}$ and $\widetilde{\H}_{0i,\mathrm{sq}}$. If we follow the same steps as before, isolating $\widetilde{\W}_{i+1+(Q_1-1)Q_2}$ instead, we would reach an expression as \eqref{eq:New_Ab3_1st_red} with one less diagonal block where each $\widetilde{\H}_m$ (still randomly chosen) has reduced the column dimension by $L$. We can thus perform these reductions inductively until we reach
\begin{equation}\label{eq:New_Ab3_final_red}
\begin{aligned}
  \scalebox{0.95}{$\begin{bmatrix}\!(\widecheck{\H}_{01}\!+\!\widecheck{\H}_{11}\!) \! & \! \cdots \! & \! (\widecheck{\H}_{0Q_2}\!+\!\widecheck{\H}_{1Q_2}\!)\!
    \end{bmatrix} \!= \!  \begin{bmatrix}\widecheck{\W}_1^{-1} \!& \! \cdots \! & \! \widecheck{\W}_{Q_2\!+\!1}^{-1}
    \end{bmatrix}$} &\\
    \times  \begin{bmatrix}
    \widecheck{\H}_{11} &  \cdots & \widecheck{\H}_{1Q_2} \\
    \widecheck{\H}_2 &  &  \\
     & \ddots &  \\
      & & \widecheck{\H}_{Q_2+1}
    \end{bmatrix}&,
\end{aligned}
\end{equation}
where $\widecheck{\H}_{0i} = \widecheck{\H}_{0}\widetilde{\vec{V}}_{i}$, with $\widecheck{\vec{V}}_{i}$ being a unitary matrix coming from a product of unitary matrices from randomly chosen blocks, $\widecheck{\H}_{1i}=[\widecheck{\H}_{1i,\mathrm{sq}} \;\; \boldsymbol{0}_{L\times (K-(Q_1+1)L)}]$ with $\widecheck{\H}_{1i,\mathrm{sq}}$ randomly chosen,  and $\widecheck{\H}_m$ for $m=0,2,\dots,Q_2+1$ are different $L\times (K-Q_1 L)$ randomly chosen blocks. It only remains to show that \eqref{eq:New_Ab3_final_red} is solvable with full-rank $\widecheck{\W}_i^{-1}$ for $i=1,\dots,Q_2+1$. If we compare \eqref{eq:New_Ab3_final_red} with \eqref{eq:New_Ab2_form} we can note that they have the same structure, but the changes in the blocks, which will allow to have $\alpha_i=1$, require a new proof.

Let us now prove that \eqref{eq:New_Ab3_final_red} allows for a solution with full-rank $\widecheck{\W}_i^{-1}$ if \eqref{eq:cond_Ab3} is fulfilled. By trivial linear algebra, we immediately note that \eqref{eq:cond_Ab3} follows from the need to have at least as many rows as columns in the matrix multiplying the RHS of \eqref{eq:New_Ab3_final_red}, since said matrix will be full-rank with probability 1. We should then check we can have full-rank $\widecheck{\W}_i^{-1}$ given \eqref{eq:cond_Ab3}. Proceeding as before we express the set of equations
\begin{equation}\label{eq:set_eq_Ab3_final_red}
    \left\{\begin{array}{l}
    \widecheck{\W}_1^{-1}\widecheck{\H}_{1i,\mathrm{sq}}+\widecheck{\W}_{i+1}^{-1}\widecheck{\H}_{i+1,\mathrm{sq}}=\widecheck{\H}_{01i,\mathrm{sq}} \\
    \widecheck{\W}_{i+1}^{-1}\widecheck{\H}_{i+1,\mathrm{r}}=\widecheck{\H}_{0i,\mathrm{r}}
    \end{array}\right., i=1,\dots,Q_2
\end{equation}
with $\widecheck{\H}_{m}=[\widecheck{\H}_{m,\mathrm{sq}} \;\; \widecheck{\H}_{m,\mathrm{r}}]$, where $\widecheck{\H}_{m,\mathrm{sq}}$ are again square, and $\widecheck{\H}_{01i,\mathrm{sq}}=\widecheck{\H}_{0i,\mathrm{sq}}+\widecheck{\H}_{1i,\mathrm{sq}}$. {Isolating $\widecheck{\W}_{i+1}^{-1}$ in the first line of \eqref{eq:set_eq_Ab3_final_red}, substituting it in the second line, and solving for $\widecheck{\W}_1^{-1}$,\footnote{{ Note that reaching from \eqref{eq:set_eq_Ab3_final_red} to \eqref{eq:W1_isol_Ab3} corresponds to the same set of steps as reaching from \eqref{eq:set_eq_Ab2} to \eqref{eq:W1_isol_Ab2} in the proof of Proposition~\ref{prop:Ab2}, with the only difference that in the first line of \eqref{eq:set_eq_Ab3_final_red} each term has an invertible (with probability 1) matrix multiplying from the right.}}} we reach
\begin{equation}\label{eq:W1_isol_Ab3}
\widecheck{\W}_1^{-1} \!\!=\! \widecheck{\vec{\mathcal{H}}}\widecheck{\vec{\mathcal{H}}}^\dagger\!\!+ \begin{bmatrix}(\vec{\Theta}_1\!-\!\widecheck{\H}_{0i,\mathrm{r}}) \!&\! \cdots \!&\! (\vec{\Theta}_{Q_2}\!-\!\widecheck{\H}_{0i,\mathrm{r}})\end{bmatrix}\widecheck{\vec{\mathcal{H}}}^\dagger\!\!+\mathbf{N}_{\widecheck{\vec{\mathcal{H}}}},
\end{equation}
where $\vec{\Theta}_m = \widecheck{\H}_{0m,\mathrm{sq}}\widecheck{\H}_{m+1,\mathrm{sq}}^{-1}\widecheck{\H}_{m+1,\mathrm{r}}$, which has dimensions $L\times (K-Q_1 L)$, $\mathbf{N}_{\widecheck{\vec{\mathcal{H}}}}$ is an $L\times L$ matrix to be selected from the left null-space of $\widecheck{\vec{\mathcal{H}}}$, 
$$\widecheck{\vec{\mathcal{H}}}=\begin{bmatrix}\widecheck{\H}_{11,\mathrm{sq}}\widecheck{\H}_{2,\mathrm{sq}}^{-1}\widecheck{\H}_{2,\mathrm{r}} \!&\cdots &\! \widecheck{\H}_{1Q_2,\mathrm{sq}}\widecheck{\H}_{Q_2\!+\!1,\mathrm{sq}}^{-1}\widecheck{\H}_{Q_2\!+\!1,\mathrm{r}}\end{bmatrix},$$
and $\widecheck{\vec{\mathcal{H}}}^\dagger$ is the left pseudo-inverse of $\widecheck{\vec{\mathcal{H}}}$. Note that the existence of said pseudo-inverse also leads to the condition \eqref{eq:cond_Ab3}. By similar arguments as in the proof of Proposition~\ref{prop:Ab2}, we have a sum between a matrix in the row space of $\widecheck{\vec{\mathcal{H}}}$, having rank $Q_2 (K-Q_1 L)$ with probability 1, and a matrix free to choose in the left null-space of $\widecheck{\vec{\mathcal{H}}}$, so $\widecheck{\W}_1$ is full-rank with probability 1 as long as $\mathbf{N}_{\widecheck{\vec{\mathcal{H}}}}$ is selected such that its rows span the whole left null-space of dimension $L-Q_2 (K-Q_1 L)$ (with probability 1). We then substitute { the expression of $\widecheck{\W}_1^{-1}$ obtained in \eqref{eq:W1_isol_Ab3} into} the first equation from \eqref{eq:set_eq_Ab3_final_red} and get
\begin{equation}\label{eq:Wi_Ab3}
\begin{aligned}
\scalebox{0.97}{$
   \widecheck{\W}_{i+1}^{-1}\widecheck{\H}_{1i,\mathrm{sq}}^{-1}\widecheck{\H}_{i+1,\mathrm{sq}}\!=\widecheck{\H}_{0i,\mathrm{sq}}\!+\!\Big(\begin{bmatrix}\widecheck{\H}_{01,\mathrm{r}} \!& \cdots &\! \widecheck{\H}_{0Q_2,\mathrm{r}}\end{bmatrix}
    \widecheck{\vec{\mathcal{H}}}^\dagger$}&\\
   \scalebox{0.97}{$ +\mathbf{I}_L-\!\widecheck{\vec{\mathcal{H}}}\widecheck{\vec{\mathcal{H}}}^\dagger-\mathbf{N}_{\widecheck{\vec{\mathcal{H}}}}-\begin{bmatrix}\vec{\Theta}_1 \!& \cdots &\! \vec{\Theta}_{Q_2}\end{bmatrix}\widecheck{\vec{\mathcal{H}}}^\dagger\Big)\widecheck{\H}_{1i,\mathrm{sq}}$}&,
\end{aligned}
\end{equation}
where we only need to check that the RHS is full-rank, since multiplying square randomly chosen blocks cannot reduce the rank (with probability 1). Reasoning as in the proof of Proposition~\ref{prop:Ab2}, $(\mathbf{I}_L-\!\widecheck{\vec{\mathcal{H}}}\widecheck{\vec{\mathcal{H}}}^\dagger)$ is a matrix in the null-space of $\widecheck{\vec{\mathcal{H}}}$ which gives rank $L-Q_2 (K-Q_1 L)$, and $\mathbf{N}_{\widecheck{\vec{\mathcal{H}}}}$ can be selected so as to not destroy said rank. Then, the other two matrices multiplying $\widecheck{\vec{\mathcal{H}}}^\dagger$ can add the remaining rank. Furthermore, $\widecheck{\H}_{0i,\mathrm{sq}}$ only shares space with $\vec{\Theta}_i$ (the rest are made of different randomly chosen blocks), so it can at most reduce rank $(K-Q_1L)$, which would be then compensated with the rank added by $\widecheck{\H}_{0i,\mathrm{r}}$, which does not share randomly chosen blocks with either $\widecheck{\H}_{0i,\mathrm{sq}}$ or $\vec{\Theta}_i$. Hence, we have proved that we can obtain full-rank $\widecheck{\W}_{i+1}^{-1}$, and this concludes the proof of Proposition~\ref{prop:Ab3}.
%We will now make use of induction to prove that \eqref{eq:set_eq_Ab3} is always solvable for full-rank $\widehat{\W}_m$ given \eqref{eq:cond_Ab3}. We start by proving the induction step. Thus, assume we can solve \eqref{eq:New_Ab3_form} with full-rank $\widehat{\W}$ for $Q_1=Q$ with $L$ and $K=K_0$ atfulfilling \eqref{eq:cond_Ab3}; we want to prove that we can also solve it for $Q_1=Q+1$ with the same $L$ and $K=K_0+L$, since then the relation \eqref{eq:cond_Ab3} is maintained. Substituting $Q_1=Q+1$ and $K=K_0+L$
\bibliographystyle{IEEEtran}
\bibliography{IEEEabrv,wax}

% \fi
\IEEEpeerreviewmaketitle

\end{document}